\documentclass[a4paper,fleqn,usenatbib]{mnras}
\usepackage{csvsimple}
\usepackage{newtxtext,newtxmath}
\usepackage[T1]{fontenc}
\usepackage{ae,aecompl}
\usepackage{graphicx}	
\usepackage{amsmath}	
\usepackage{amssymb}	

\newcommand{\msun}{\mbox{$M_\odot$}} 

\newcommand{\rsun}{\mbox{$R_\odot$}}
\newcommand{\mstar}{M$_{\star}$}

\newcommand{\rstar}{R$_{\star}$}
\newcommand{\mjup}{M$_{\rm J}$}
\newcommand{\rjup}{R$_{\rm J}$}

\newcommand{\kms}{\mbox{km s$^{-1}$}}

\title[Neutral species in WASP-121b]{Detection of neutral atomic species in the ultra-hot jupiter WASP-121b}

\author[S. H. C. Cabot et al.]{
Samuel H. C. Cabot$^{1}$\thanks{E-mail: sam.cabot@yale.edu},
Nikku Madhusudhan$^{2}$,
Luis Welbanks$^{2}$,
Anjali Piette$^{2}$,\newauthor
and Siddharth Gandhi$^{3}$
\\
$^{1}$Yale University, 52 Hillhouse, New Haven, CT 06511, USA\\
$^{2}$Institute of Astronomy, Madingley Road, Cambridge CB3 0HA, UK\\
$^{3}$Department of Physics, University of Warwick, Coventry CV4 7AL, UK
}

\date{Accepted XXX. Received YYY; in original form 2019 November 18}

\pubyear{2019}

\begin{document}
\label{firstpage}
\pagerange{\pageref{firstpage}--\pageref{lastpage}}
\maketitle

\begin{abstract}
The class of ultra-hot Jupiters comprises giant exoplanets undergoing intense irradiation from their host stars. They have proved to be a particularly interesting population for their orbital and atmospheric properties. One such planet, WASP-121~b, is in a highly misaligned orbit close to its Roche limit, and its atmosphere exhibits a thermal inversion. These properties make WASP-121~b an interesting target for additional atmospheric characterization. In this paper, we present analysis of archival high-resolution optical spectra obtained during transits of WASP-121~b. Artifacts from the Rossiter-McLaughlin effect and Center-to-Limb Variation are deemed negligible. However, we discuss scenarios where these effects warrant more careful treatment by modeling the WASP-121 system and varying its properties. We report a new detection of atmospheric absorption from H$\alpha$ in the planet with a transit depth of $1.87\pm0.11\%$. We further confirm a previous detection of the Na I doublet, and report a new detection of Fe I via cross-correlation with a model template. We attribute the  H$\alpha$ absorption to an extended Hydrogen atmosphere, potentially undergoing escape, and the Fe I to equilibrium chemistry at the planetary photosphere. These detections help to constrain the composition and chemical processes in the atmosphere of WASP-121~b.
\end{abstract}

\begin{keywords}
planets and satellites: atmospheres
\end{keywords}

\section{Introduction} \label{sec:intro}

Ultra hot Jupiters (UHJs) display some of the most extreme physics of all known exoplanets. These massive gas giants have close-in, often tidally locked orbits with their host stars, and are subject to extremely strong irradiation. UHJs offer fascinating case studies in their own right: they have equilibrium temperatures in excess of 2000 K \citep{Fortney2008, Parmentier2018}, can contain vaporized metals in their atmospheres \citep{Hoeijmakers2018, Casasayas2018}, and often have orbits completely misaligned from their host star's rotation \citep{Triaud2010, Anderson2018}. Some orbit close to their Roche limit, and are on the verge of tidal disruption \citep{Delrez2016}. Others have extended, escaping atmospheres \citep{Yan2018}. These unusual properties test and inform theories of how hot Jupiters form and evolve, and improve our understanding of their atmospheric chemistry and dynamics. 

An outstanding problem is how these planets attain their close-in orbits. An {\it in-situ} scenario is often thought unlikely since proximity to the host star is not conducive for gas giant formation, though recent studies have shown rapid accretion of hot gas may be possible \citep{Batygin2016}. The other scenario is migration, either through the protoplanetary disk, or via dynamical scattering with other bodies \citep{Dawson2018}. Chemical tracers such as relative amounts of C and O help constrain whether formation occurs within or beyond the snow line \citep{oberg2011,madhu2014c}. Present day orbital obliquities favor dynamical scattering \citep{Triaud2010}, as does the fact that most semimajor axes are lower bounded by twice the Roche limit ($a/a_R \sim 2$) \citep{Ford2006}. However, several hot Jupiters in orbits of $a/a_R < 2$ require alternative or more complex histories, such as tidal decay \citep[see][for more detailed discussions]{Delrez2016, Dawson2018}. A comprehensive understanding of the dynamics and chemistry of hot Jupiters is thus imperative to resolve these issues. 

With transit spectroscopy, UHJs are some of the easiest planets to characterize. High temperatures give UHJs relatively high day-side flux contrasts with respect to their host stars, and their large radii (usually 1-2 \rjup) yield strong transit depths. To date, several UHJs have been studied through transmission and emission spectroscopy. Some examples include KELT-9 b \citep{Hoeijmakers2018}, MASCARA-2 b \citep{Casasayas2019}, WASP-121~b \citep{Evans2017}, WASP-33 b \citep{Nugroho2017}, WASP-103 b \citep{Cartier2017}, and WASP-18 b \citep{Sheppard2017, Arcangeli2018, Espinoza2019}. A number of neutral and ionized atomic species have been predicted in the atmospheres of such UHJs \citep[][]{Kitzmann2018,Lothringer2018}. In particular, phase-resolved high-resolution transmission spectroscopy has proved an excellent way to probe hot Jupiter atmospheres. The cross-correlation approach \citep{Snellen2010} involves comparing observed spectra with a model template of a species in order to detect forests of weak absorption lines. Molecular, atomic and ionized species have been detected this way \citep{Snellen2010, birkby2018, Alonso-Floriano2019, Hoeijmakers2019}. Strong features, such as the Na doublet and Balmer lines can be directly recovered by co-adding multiple in-transit spectra \citep{Wyttenbach2015, Casasayas2019}. With sufficient phase-coverage, it is possible to resolve day-to-night side winds \citep{Louden2015} and extended or escaping atmospheres \citep{Ehrenreich2015}.

One planet of particular interest is the ultra hot Jupiter WASP-121~b ($T_{\rm eq} = 2358 \pm 52$ K), which is in a near-polar orbit around a bright ($V=10.44$) F-type star \citep{Delrez2016}. It's semi-major axis is only $\sim 1.15$ times its Roche limit, suggesting the planet is on the verge of tidal disruption. Deformation models suggest the planet may have radius $R_{\rm sub} \sim 2 R_{\rm jup}$ at its sub-stellar point \citep{Delrez2016}. The bright host star and extended atmosphere makes the planet a prime target for characterization. Indeed, the planet has been studied extensively. \citet{Evans2016} detected H$_2$O in the atmosphere of WASP-121~b using a transmission spectrum obtained using {\it the Hubble Space Telescope} (HST) WFC3 spectrograph and ground-based observations. \cite{Evans2017} reported a detection of H$_2$O and a thermal inversion in the dayside atmosphere using a thermal emission spectrum obtained with HST WFC3. However, a direct detection of TiO or VO proved elusive in subsequent transit and secondary eclipse studies \citep{Evans2018, Evans2019}. Recent optical phase curves from {\it TESS} along with other data confirm the presence of a thermal inversion in the dayside atmosphere \citep{Daylan2019, Bourrier2019}. While species such as H-, TiO, and VO have been suggested as possible inversion-causing absorbers in the planet \citep{Daylan2019, Bourrier2019}, a variety of other absorbers may also be responsible \citep{Molliere2015,Gandhi2019}.  Separately, \citet{Salz2019} suggested excess broadband NUV absorption might be due to Fe II, a species later detected by \citet{Sing2019}, in addition to Mg II. The ionized gas extends out to $R_p/R_s \sim 0.3$, and might be undergoing atmospheric escape or be confined due to a magnetic field.

WASP-121~b currently lacks a comprehensive transit study at high-resolution in the optical regime. Here, we analyze three transits of WASP-121~b observed by HARPS. While \citet{Sindel2018} detect the Na doublet in transmission using one order of this dataset, we present additional detections of the H$\alpha$ line and Fe I by analyzing the full wavelength coverage. Our paper is organized as follows. In \S\ref{sec:obs} and \S\ref{sec:methods} we present the dataset, preprocessing steps, and review methodology of phase-resolved high-resolution transmission spectroscopy. We present atomic detections in \S\ref{sec:res}, and discuss their implications for the atmosphere of WASP-121~b in \S\ref{sec:disc}. Additionally in \S\ref{sec:disc}, we investigate the impact of the Rossiter-McLaughlin effect and Center-to-Limb Variation on the transmission spectra.

\section{Observations}
\label{sec:obs}

Our dataset consists of archival optical spectra of WASP-121 acquired by the HARPS (High-Accuracy Radial-velocity Planet Searcher) echelle spectrograph, located at the ESO La Silla 3.6m telescope. At a resolution of $\sim$115,000, HARPS coverage spans 380-690nm over 68 spectral orders. The raw data were reduced with the HARPS Data Reduction Software (DRS) v3.8, which performs blaze correction and Th-Ar wavelength calibration, and produces one-dimensional spectra rebinned onto a uniform 0.01\AA\, barycentric rest-frame wavelength grid. Observations were conducted over the nights of 31 December 2017, 09 January 2018, and 14 January 2018 (hereafter Nights 1, 2, and 3) as part of the Hot Exoplanet Atmospheres Resolved with Transit Spectroscopy program (HEARTS) Program 0100.C-0750(C). There are 140 exposures in total, with 49 acquired during the transit (Table~\ref{tab:exp}).

\begin{figure*}
   \centering
    \includegraphics[width=\linewidth,angle=0,trim={0cm 0cm 0cm 0cm},clip]{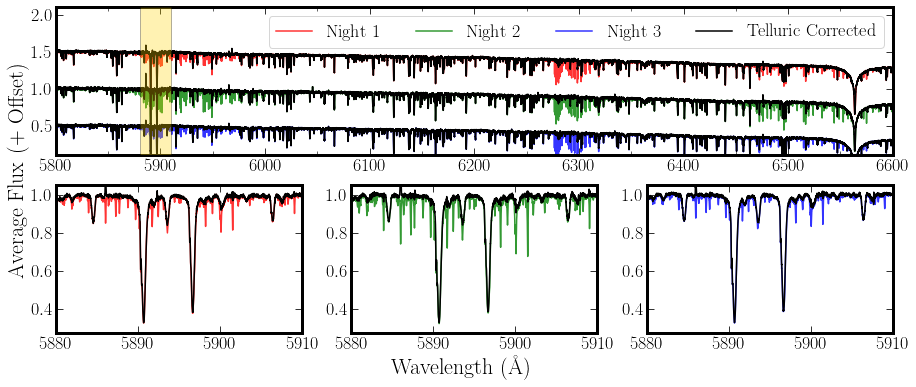}
   \caption{
      Telluric correction with \texttt{Molecfit}. {\it Top panel}: Co-added and normalized spectra for Nights 1, 2 and 3 are depicted in color (red, green and blue respectively). The black curves are the co-added spectra after dividing out the telluric model. The yellow stripe denotes the zoom-in wavelength range. {\it Bottom panels}: same as above, but cropped in a narrow wavelength range around the Sodium Doublet, for each night. Water vapour is the predominant telluric absorber in this range.
      }
    \label{fig:tel_cor}
\end{figure*}

Absorption due to H$_2$O and O$_2$ in Earth's atmosphere produces strong telluric features at wavelengths upwards of $\sim$500nm. We use the ESO tool \texttt{molecfit} v1.5.7, which fits a line-by-line radiative transfer model (LBLRTM) of Earth's transmission spectrum to observed telluric features \citep{smette2015}. Whereas some previous works construct an empirical telluric model \citep{Wyttenbach2015, Casasayas2018}, we opt to use \texttt{molecfit} for the much lower S/N exposures analyzed here. However, care must be taken to avoid fitting to stellar absorption lines \citep{Allart2017, Hoeijmakers2019, Casasayas2019}. For each exposure, on each night, we shift the DRS wavelength solution to the telescope rest-frame using the Barycentric Earth Radial Velocity (BERV). We identify regions suitable for telluric fitting by: 1) obtaining centroids of $\sim$300 of the strongest known telluric features in a model telluric spectrum obtained from \texttt{molecfit} \citep{smette2015, Allart2017}; 2) obtaining a $T_{eff} = 6500$ K \texttt{PHOENIX} stellar model \citep{Husser2013}; 3) Doppler shifting the model to the same radial velocity as WASP-121 (the sum of the systemic velocity and BERV); 4) choosing tellurics preselected in step (1) whose locations fall outside of features in the stellar model; and 5) selecting small wavelength ranges centered on these telluric features. We provide \texttt{molecfit} location and ambient weather parameters from the HARPS DRS output, and fit parameters similar to those used by \citet{Allart2017}. After fitting, we use the included \texttt{calctrans} tool to calculate a high-resolution model telluric spectrum over the full wavelength range. We finally divide the observed spectrum by the model (Figure \ref{fig:tel_cor}). There is slight over-correction at 5885 \AA, possibly from an extra water feature in the line-database; however it is offset sufficiently as to not influence the planetary Na absorption. There is no apparent telluric sodium emission, which we verified by inspecting simultaneous sky spectra from fiber B.

\setlength{\tabcolsep}{5pt}
\renewcommand{\arraystretch}{0.8}
\begin{table*}
\begin{tabular}{l | c | c | c }
Night                 & 2017-12-31 & 2018-01-09 & 2018-01-14\\
\hline
\hline
$T_{\rm start}$ (UTC) & 01:39      & 00:34      & 01:18     \\
$T_{\rm end  }$ (UTC) & 08:16      & 08:38      & 08:43     \\
$t_{\rm exp}$ (s)     & 570-720        & 500-600        & 500-660       \\
SNR$_{\rm cont}$      & $\sim$36   & $\sim$25   & $\sim$31 \\
$N_{\rm in-transit}/N_{\rm total}$ & 13/35 & 18/55 & 18/50  \\
sec $z$     &1.339 $\rightarrow$ 1.015 $\rightarrow$ 1.329  & 1.501 $\rightarrow$ 1.015 $\rightarrow$ 1.626 & 1.225 $\rightarrow$ 1.015 $\rightarrow$ 1.826  \\
$\phi$      &0.883 $\rightarrow$ 0.099 & 0.907 $\rightarrow$ 0.170 & 0.853 $\rightarrow$ 0.095\\
\end{tabular}
\caption{Exposure information for each night of observation. Rows correspond to 1) observing start time; 2) observing end time; 3) min. and max. exposure durations; 4) approximate signal-to-noise of wavelength bins along the continuum; 5) number of in-transit exposures to number of total exposures; 6) airmass evolution throughout the night; 7) WASP-121~b orbital phase coverage (mid-transit at $\phi = 0$ and 1).}
\label{tab:exp}
\end{table*}

\setlength{\tabcolsep}{3pt}
\renewcommand{\arraystretch}{1.0}
\begin{table}
\begin{tabular}{l l l | l}
Parameter                & Symbol & Unit & Value \\
\hline
\hline
{\bf Star} & & & \\
\hline
V-magnitude              & $m_V$           & mag.            & 10.4   \\
Effective Temperature    & $T_{eff}$       & K               & 6460               $^{+140       }_{-140       }$ \\
Proj. Rot. Velocity      & $v {\rm sin} i$ & \kms            & 13.56              $^{+0.69      }_{-0.68      }$ \\
Stellar Mass             & \mstar          & \msun           & 1.353              $^{+0.080     }_{-0.079     }$ \\
Stellar Radius           & \rstar          & \rsun           & 1.458              $^{+0.030     }_{-0.030     }$ \\
Stellar Semi-Amplitude   & $K_*$           & \kms            & 181                $^{+6.3       }_{-6.4       }$ \\
\hline
{\bf System} & & & \\
\hline
Mid-Transit Time         & $T_0$           & HJD$_{\rm TDB}$ & 2456635.70832      $^{+0.00011   }_{-0.00010   }$ \\
Transit Duration         & $\tau$          & d               & 0.1203             $^{+0.0003    }_{-0.0003    }$ \\
Period                   & $P$             & d               & 1.2749255          $^{+0.00000020}_{-0.00000025}$ \\
Semi-major Axis          & $a$             & A.U.            & 0.02544            $^{+0.00049   }_{-0.00050   }$ \\
Limb-darkening Coefficient  & $u_{1,r'}$   & -               & 0.290              $^{+0.014     }_{-0.014     }$ \\
Limb-darkening Coefficient  & $u_{2,r'}$   & -               & 0.305              $^{+0.325     }_{-0.007     }$ \\
Systemic Velocity        & $\gamma$        & \kms            & 38.350             $^{+0.021     }_{-0.021     }$ \\
\hline
{\bf Planet} & & & \\
\hline
Orbital Inclination      & $i_p$           & deg.            & 87.6               $^{+0.6       }_{-0.6       }$ \\
Sky Proj. Obliquity      & $\beta$         & deg.            & 257.8              $^{+5.3       }_{-5.5       }$ \\
Equilibrium Temperature  & $T_{eq}$        & K               & 2358               $^{+52        }_{-52        }$ \\
Planetary Mass           & M$_p$           & \mjup           & 1.183              $^{+0.064     }_{-0.062     }$ \\
Planetary Radius$^*$         & R$_p$           & \rjup           & 1.865              $^{+0.044     }_{-0.044     }$ \\
Planetary Semi-Amplitude$^{**}$  & $K_p$           & \kms            & 217               $^{+19       }_{-19       }$ \\
\\
\end{tabular}
\caption{Literature values for stellar, orbital, and planetary parameters for WASP-121~b and its host star.}
\footnotesize{All values from \citet{Delrez2016}. \\ $^*$corrected for asphericity.\\
$^{**}$calculated from $M_p/M_* = K_*/K_p$.}
\label{tab:param}
\end{table}

\section{Methods}
\label{sec:methods}

In this section, we discuss our extraction of the planetary transmission spectrum using the technique of \citet{Wyttenbach2015}. We treat each night separately throughout the analysis, and propagate the Poisson-uncertainties of the data. We also discuss our cross-correlation procedure, based on \citet{Snellen2010}, which lends itself to detecting species with a multitude of weak features.

\subsection{Transmission Spectra}

During a transit event, an exoplanet blocks part of its host star along our line of sight and causes the star to appear dimmer. The change in brightness is approximately the ratio of the area occulted by the planet to the area of the stellar disk,
\\
\begin{equation}
\Delta_0^2 = \Big(\frac{R_{p}}{R_*}\Big)^2.
\end{equation}
\\
This wavelength-integrated quantity is often referred to as the white-light transit depth. Beyond the gray photospheric radius, the planet's atmosphere absorbs additional light at specific wavelengths, which produces its transmission spectrum. The amount of absorption at a given wavelength depends on the abundance and cross-section of the absorbing species.

Given an atmospheric height $H(\lambda)$, we denote the wavelength-dependent transit depth as:
\\
\begin{equation}
\Delta_\lambda^2 = \Big(\frac{R_p + H(\lambda)}{R_*}\Big)^2 = \Big(\frac{R_p}{R_*}\Big)^2 + \Big(\frac{H(\lambda)}{R_*}\Big)^2 + \frac{2R_pH(\lambda)}{R_*^2} \simeq \Delta_0^2 + \frac{2R_pH(\lambda)}{R_*^2}
\end{equation}
\\
where we assume $H(\lambda) << R_*$ (an atmosphere generally extends 5-10 scale heights, or several thousand kilometers for a hot-Jupiter) \citep{madhu2014a}. Using the ephemeris of \citet{Delrez2016} (Table~\ref{tab:param}), we identify 13, 18, and 18 in-transit exposures for Nights 1, 2 and 3 respectively. We denote in-transit and out-of-transit spectra as $f(\lambda, t_{\rm in})$, and $f(\lambda, t_{\rm out})$. Each night's time-series covers the entire $\sim$2.5 hour transit, plus several hours of baseline exposures before and after the transit. 

Several effects can affect the fidelity of the transmission spectrum. These include: telluric absorption by species in Earth's atmosphere; the reflex motion of the host star induced by the planet's orbit; the Rossiter-Mclaughlin (RM) effect; Center-to-Limb Variation (CLV); and the planet's changing radial velocity throughout the transit. All of these are resolved at the HARPS spectral resolution \citep{Wyttenbach2015, Louden2015, Allart2017, Yan2017, Casasayas2019}. We systematically address each of these effects. Having corrected for tellurics with \texttt{molecfit}, we linearly interpolate all spectra onto a common wavelength grid, and Doppler shift each to correct for the stellar reflex velocity:
\\
\begin{equation}
    v_{\rm reflex} = -K_*\sin{2\pi\phi}
\end{equation}
\\
Note, there is no apparent Interstellar Medium (ISM) Sodium absorption. Otherwise, the reflex velocity correction could prevent its cancellation when we later divide in-transit and out-of-transit spectra \citep{Casasayas2018}. Following a thorough investigation (Section \ref{sec:clv}), we find CLV and RM induced effects lie at the noise level of the data, and we neglect their correction. Since WASP-121~b is in a near-polar orbit ($\beta \sim 257.8^\circ$), it obscures regions of similar velocity throughout the transit. As such, the RM distortion of stellar absorption lines blurs out when stacked in the planet rest-frame. Also, WASP-121 is a hot, F6V-type star, and the CLV effect on the transmission spectrum is expected to be negligible \citep{Yan2017}.

\begin{figure*}
   \centering
    \includegraphics[width=\linewidth,angle=0,trim={0cm 0cm 0cm 0cm},clip]{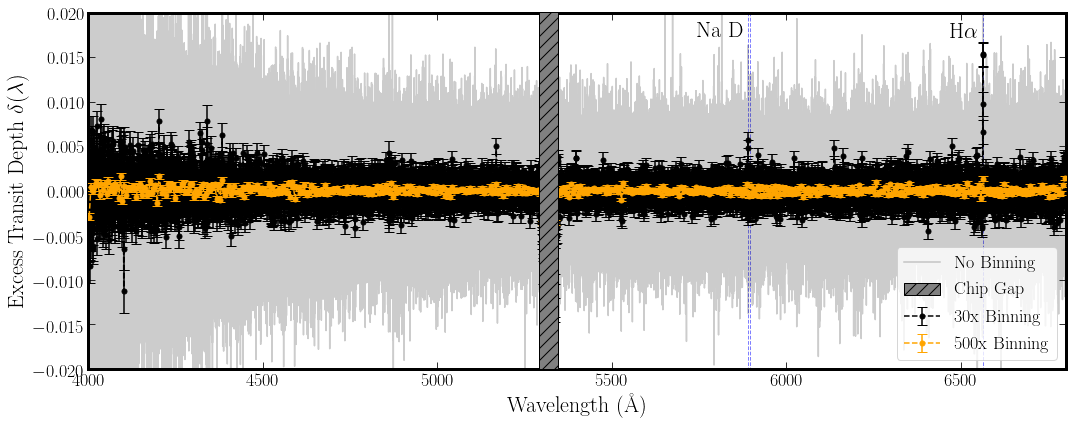}
      \caption{
      Excess transit depth from the atmosphere of WASP-121~b, over the full analysed wavelength range. Particularly notable are the Sodium D-lines and H$\alpha$ feature; although a few other features visibly extend from the continuum. We show the unbinned data, and data binned by 30x and 500x pixels. The HARPS chip gap is masked in our analysis.
      }
    \label{fig:tr_full}
\end{figure*}

We create a master out-of-transit spectrum by co-adding individual out-of-transit spectra:
\\
\begin{equation}
    \hat{f}_{\rm out}(\lambda) = \sum_{t_{\rm out}}{f(\lambda, t_{\rm out})}
\end{equation}
\\
and compute individual transmission spectra as:
\\
\begin{equation}
    \mathfrak{R}(\lambda, t_{\rm in}) = \frac{f(\lambda, t_{\rm in})}{\hat{f}_{\rm out}(\lambda)}
\end{equation}
The continuum level of each spectrum it is affected by throughput variations from the instrument and weather. Therefore we normalize each $\mathfrak{R}(\lambda, t_{\rm in})$ by fitting and dividing by a 5$^{\rm th}$-order polynomial. The planet's apparent radial velocity is given by,
\\
\begin{equation}
    v_{\rm pl} = K_p\sin{2\pi\phi} + \gamma
    \label{eqn:vpl}
\end{equation}
\\
where $K_p$ is the semi-amplitude and $\gamma$ is the systemic velocity (Table \ref{tab:param}). Throughout the transit, $v_{\rm pl}$ changes by $\sim100$ \kms, corresponding to a Doppler shift of 2 \AA, or 200 pixels. To avoid smearing out the atmospheric signal, we Doppler shift each $\mathfrak{R}(\lambda, t_{\rm in})$ by -$v_{\rm pl}(t_{\rm in})$ and stack them in the rest-frame of the planet:
\\
\begin{equation}
\hat{\mathfrak{R}}(\lambda) = \sum_{t_{\rm in}} \mathfrak{R}(\lambda, t_{\rm in})|_{v_{\rm pl}(t_{\rm in})}
\end{equation}
\\
Finally, we apply a median filter of width 1501 pixels to remove remaining broadband variations. It is important to precisely define $\hat{\mathfrak{R}}(\lambda)$. Since we have lost continuum information, $\hat{\mathfrak{R}}(\lambda)$ corresponds to the transmission spectrum of the planet, after removing the white-light transit depth $\Delta_0^2 = (R_p/R_*)^2$. Values less than unity correspond to absorption by the planet's atmosphere. For further analysis, we define the quantity:
\\
\begin{equation}
\delta(\lambda) \equiv -\hat{\mathfrak{R}}(\lambda)+1 = \Delta_\lambda^2 - \Delta_0^2 \simeq \frac{2R_pH(\lambda)}{R_*^2}
\end{equation}
\\
where $\delta(\lambda)$ is the {\it excess} transit depth caused by the atmosphere of the planet. Positive values correspond to atmospheric absorption.

Normalization with a polynomial is common practice in previous literature  \citep[e.g.][]{Seidel2019, Casasayas2018}, and can be applied before division by the master-out \citep{Allart2017}, after division \citep{Seidel2019}, or after stacking individual transmission spectra \citep{Casasayas2018}. Typically it is done with a 3$^{\rm rd}$ or 4$^{\rm th}$ degree polynomial; we found a 5$^{\rm th}$ degree suitable for the large wavelength range in our analysis. Finally, we stack the co-added transmission spectra from each night to obtain a master transmission spectrum (Figure \ref{fig:tr_full}).

\subsection{Cross-Correlation}

Strong features such as the Na doublet, H Balmer lines, the Ca II triplet, the Mg I triplet, and He I have been detected in hot gas giants by directly extracting their transmission spectra \citep{Wyttenbach2015, Casasayas2018, Nortmann2018, Cauley2019}. However, atomic and molecular species can produce a dense forest of thousands of weak absorption lines \citep{Hoeijmakers2019, Gandhi2019b}. We can search for these species by cross-correlating with a model transmission spectrum, which stacks the signal from all of the absorption lines. This approach has been used successfully in the optical regime \citep{Nugroho2017, Hoeijmakers2018, Hoeijmakers2019}, as well as the near-infrared \citep{Snellen2010, Brogi2012, rodler2013, lockwood2014, Birkby2013, piskorz2016, Birkby2017, brogi2018, Hawker2018}. 
We use the \texttt{X-COR} pipeline, which was previously used to detect CO, H$_2$O, and HCN in the dayside atmospheres of hot Jupiters \citep{Hawker2018, Cabot2019}. In the near-infrared, strong telluric absorption warrants aggressive preprocessing, often through use of Principal-Component-Analysis (PCA) or its uncertainty weighted version (SYSREM) \citep{tamuz_2005}. In the optical regime, the \texttt{molecfit} model is sufficient for telluric correction. We apply a sliding filter to each spectrum which flags $\geq5\sigma$ outliers and replaces them with the median value in the window. We mask the chip gap and $1\%$ of data from either end of the full spectrum which suffer from low throughput or strong telluric contamination. Following from the previous section, all residuals are currently in the planetary rest-frame. Finally, we remove any remaining broadband variations by applying a 75-pixel width high-pass filter, and subtracting the mean of each wavelength bin.

Cross-correlation involves a model template, which is derived from a theoretical transmission spectrum of one or multiple species. Model spectra generation is discussed in the following section. To obtain the template, we subtract the maximum value in a  0.008\AA-sliding window across the model spectrum to remove its continuum. The template is convolved with a narrow Gaussian filter of FWHM $= 0.8$ \kms\ to match approximately the wavelength sampling of the HARPS detector \citep{Hoeijmakers2019}, and subsequently normalized to unity. We define our cross-correlation-function (CCF) as a function of velocity and time:

\begin{equation}
{\rm CCF}(v, t) = \frac{\sum_{i}m_i|_{v}w_ix_i(t)}{\sum_{i}m_i|_{v}w_i}
\label{eqn:ccf}
\end{equation}
\\
where $m_i$ is our model template Doppler shifted by $v$, $x_i(t)$ is the observed transmission spectrum over wavelength bins $i$, and $w_i$ are weights assigned to each bin. Weights are the time-axis variance of each wavelength bin, which effectively down-weights noisy pixels that are affected by low throughput, or lie in the cores of telluric or stellar lines \citep{Brogi2016}. The normalisation term in Equation \ref{eqn:ccf} preserves the intrinsic strength of the absorption features. That is, the CCF returns a weighted average of line-depths in the data \citep{Hoeijmakers2019}. We perform cross-correlation over a velocity grid spanning $-600 \leq v \leq 600$ \kms\ in steps of 2.0 \kms. 

\begin{figure*}
  \centering
   \includegraphics[width=\linewidth,angle=0,trim={0cm 0cm 0cm 0cm},clip]{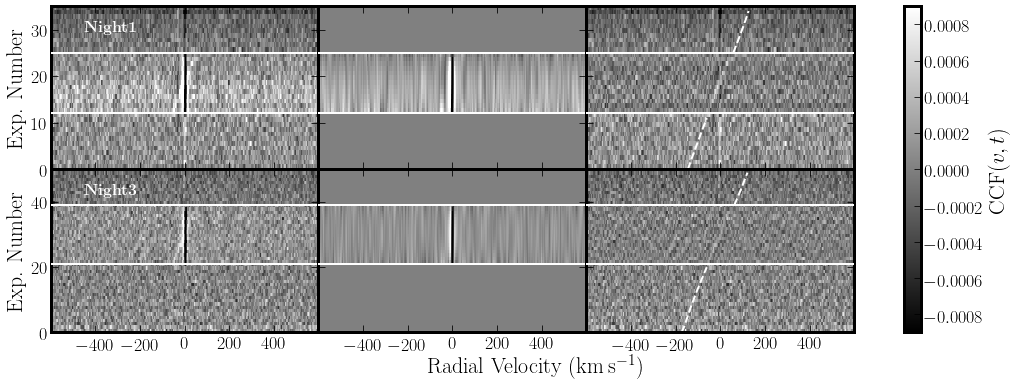}
      \caption{
      Time-series cross-correlation functions (CCFs) of WASP-121 spectra with a \texttt{PHOENIX} model template. Spectra were Doppler shifted by the known $V_{\rm sys}$ of WASP-121 prior to cross-correlation. Horizontal white lines mark the start and end of transit. Each row shows the procedure for a different night of observation. {\it First Column}: Time-series CCFs. {\it Second Column}: Polynomial fit to each column of the CCFs, which serves as a model of the Doppler Shadow RM residual. {\it Third Column}: CCFs after subtracting the Doppler Shadow Model. White dashed lines mark the velocity of WASP-121~b before and after transit. A faint white trail in the in-transit frames represents absorption by the atmosphere of WASP-121~b through features in common with the \texttt{PHOENIX} model.
      }
    \label{fig:ccf_pho}
\end{figure*}

If the model template contains species native to the host star, then this procedure reveals the Doppler Shadow from the Rossiter-McLaughlin effect \citep{Cegla2016}. Indeed, the Doppler Shadow dominates the CCF, and must be removed for atmospheric analysis. We cross-correlate with a $T_{eff} = 6500$ K \texttt{PHOENIX} stellar template. Then the Doppler Shadow and other broadband variations are modelled by fitting a 3$^{\rm rd}$-order polynomial to the in-transit CCF values at each sampled velocity. Our approach differs from that used in a previous study of KELT-9 b \citep{Hoeijmakers2019}, where a time-varying Gaussian profile is iteratively fit to the shadow and atmospheric signal. We attempted this approach but could not obtain robust fits. The WASP-121 RM residual has asymmetric negative wings on its edges, and is not well-approximated by a Gaussian. The CCF RM residual is a reflection of the RM residuals of individual lines, which also have this shape (see section \ref{sec:clv}). The shape might come from a combination of the weak Center-to-Limb Variation and normalization step; in-transit stellar lines should be `missing' the flux occulted by the planet, but these might cause the appearance of excess flux in the wings. We attempted simultaneous fitting of a positive and negative Gaussian profile, but this allowed too many free parameters for the relatively low S/N data (WASP-121 is about three magnitudes fainter than KELT-9, and we have fewer in-transit exposures per night). Our polynomial-fit approach works well in our case because the RM effect spans a small velocity range, whereas the planetary signal is spread over $\sim 100$ \kms. When cross-correlating with other templates, we scale the Doppler Shadow model to fit the CCF, and subsequently subtract it to isolate the atmospheric signal. The full process is shown in Figure \ref{fig:ccf_pho}. We excluded Night 2 from cross-correlation analysis due to its low S/N.

Since the planet's radial velocity changes throughout the transit, the absorption signal appears as a moving trail. The planet velocity is uniquely determined by a certain combination of semi-amplitude and systemic velocity (Equation \ref{eqn:vpl}). We sample these two parameters from a grid, and for each combination, Doppler shift the CCFs by the corresponding planet velocities, and subsequently co-add them. The correct combination stacks the planet signal in-phase, boosting the S/N. Incorrect combinations gives us a noise estimate from which we compute detection significance.

\subsection{Model Spectra}
\label{sec:mod}

We model high-resolution spectra for WASP-121~b assuming a H$_2$-He dominated clear atmosphere with gaseous atomic Fe. The model spectra are generated using an adaptation of the AURA model for exoplanetary transmission spectra \citep[e.g.,][]{Pinhas2018}. The spectra are computed using line-by-line radiative transfer in a plane parallel atmosphere in transmission geometry. The model atmosphere is divided in 100 layers uniformly distributed in log space between pressures of $10^{2}-10^{-6}$ bar and assumes hydrostatic equilibrium along with a uniform chemical volume mixing ratio of the species of interest. The model sets the planetary radius, uncorrected for asphericity, at a reference pressure of 100 mbar.  We adopt an isothermal temperature profile at 2400 K, roughly the equilibrium temperature of the planet assuming full redistribution and zero albedo. The model spectra are calculated using 3$\times$10$^5$ wavelength points in a uniform wavelength grid from 0.4-0.7 $\mu$m, corresponding to a resolution of R$\sim$4-7$\times$10$^5$, higher than the resolving power of the instrument. The system parameters (e.g. planetary and stellar radii, planetary gravity, and orbital semimajor axis) are obtained from \citet{Delrez2016} (see Table \ref{tab:param}).

Motivated by recent elemental detections in ultra-hot Jupiters we search for multiple species in the atmosphere of WASP-121b and empirically detect Fe. Our present model atmosphere considers absorption due to gaseous atomic Fe along with the effects of collision-induced-absorption (CIA) due to H$_2$-H$_2$ and H$_2$-He \citep{Richard2012} and H$_2$ Rayleigh scattering. The atomic opacity is calculated following the methods of \citet{gandhi_2017} with absorption cross sections computed from NIST \citep{Kramida2018}. We assume the volume mixing ratio of Fe in the atmosphere to be 10$^{-7}$, which is a lower limit on the Fe abundance feasible in this atmosphere as discussed below. The volume mixing ratios of H and He are calculated assuming a He/H$_2$ ratio of 0.17 and requiring the sum of the abundances to be unity. 

\begin{figure}
    \centering
     \includegraphics[width=0.5\textwidth]{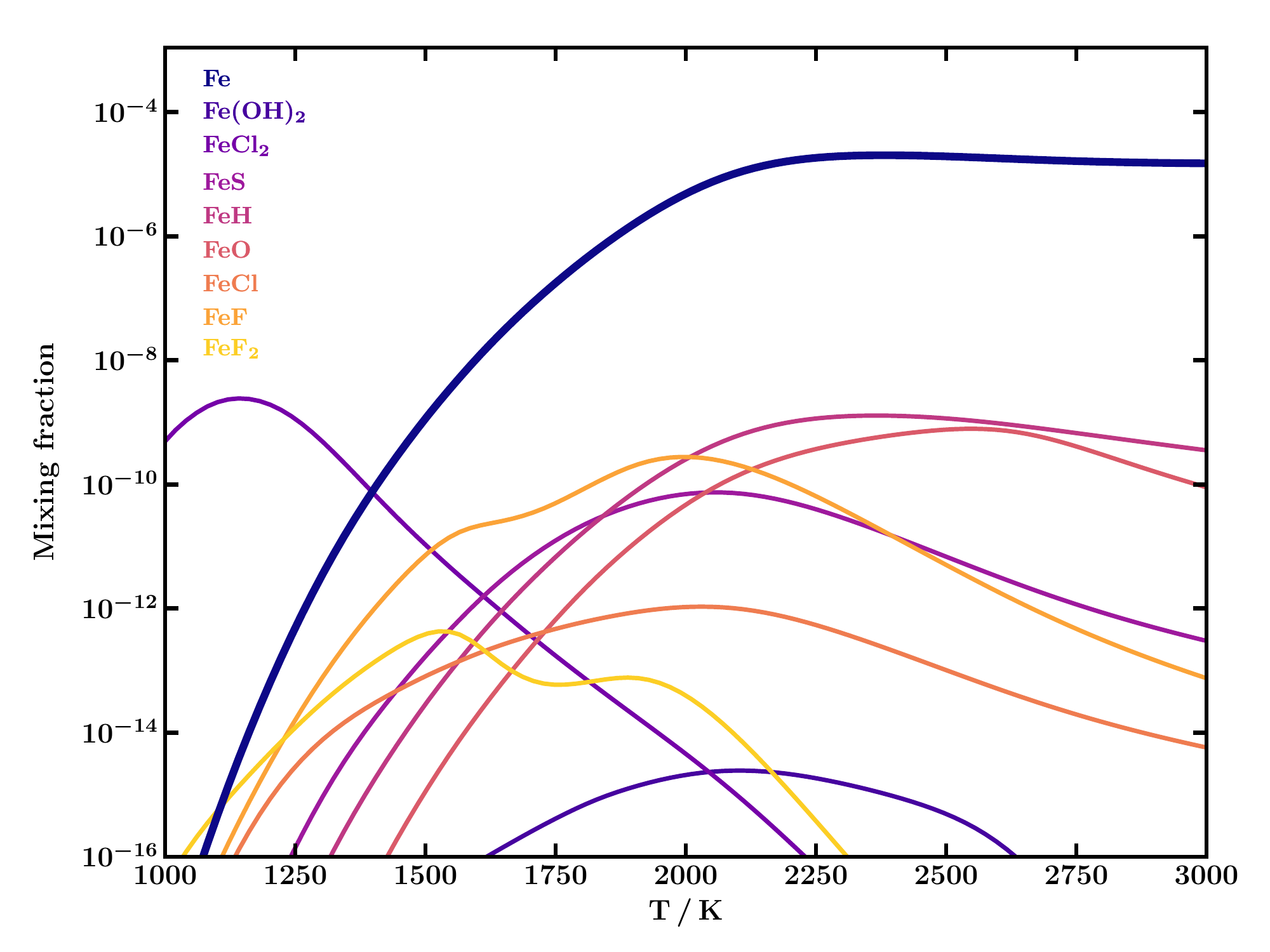}
    \caption{Equilibrium chemical abundances of several Fe-based neutral gaseous species at 1 mbar as a function of temperature. Abundances are calculated using the \textsc{HSC Chemistry} software (version 8) assuming solar elemental abundances \citep{Asplund2009}. At the equilibrium temperature of WASP-121~b ($\sim$2400 K), neutral atomic Fe (bold line) is the dominant species.}
    \label{fig:eqmfe}
\end{figure}

We verify that neutral atomic Fe is the dominant Fe species using equilibrium chemistry calculations computed with the \textsc{HSC Chemistry} software (version 8) \citep[e.g.][]{pasek2005,bond2010,elser2012, madhu2012,moriarty2014,harrison2018}. These calculations assume solar elemental abundances \citep{Asplund2009} and include the same species as \citet{harrison2018} plus gaseous, solid, neutral and ionic molecules and atomic forms of Fe, Ti, V, Cr and Mg. Figure \ref{fig:eqmfe} shows the equilibrium abundances of several neutral gaseous Fe species between 1000-3000 K at a pressure of 1 mbar nominally corresponding to the optical photosphere. At the equilibrium temperature of WASP-121~b ($\sim$2400 K), neutral atomic Fe is the dominant form of Fe and should therefore be the most easily detected \citep[also see][]{Kitzmann2018,Lothringer2018}. We note that the terminator temperature probed in the optical at high resolution, i.e., going as high up in the atmosphere as 1 mbar - 1 $\mu$bar, can be significantly lower than the equilibrium temperature. As such the Fe abundance can be significantly lower than the maximum Fe abundance possible at high T of $\sim$10$^{-5}$, as seen in Figure~\ref{fig:eqmfe}. 

\section{Results}
\label{sec:res}

\subsection{Transmission Spectrum of WASP-121~b}

\begin{figure*}
   \centering
    \includegraphics[width=\linewidth,angle=0,trim={0cm 0cm 0cm 0cm},clip]{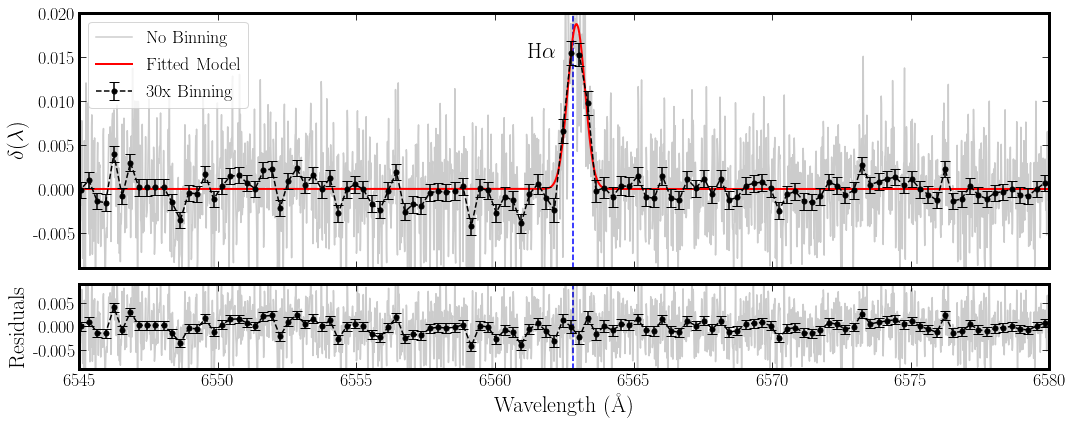}
      \caption{
      Excess transit depth from WASP-121~b for the H$\alpha$ line. {\it Top Panel}: combined data from all three nights of observation, stacked in the rest-frame of the planet. Also shown are the data binned 30x, and the best-fit Gaussian profile to the absorption feature. {\it Bottom Panel}: residuals in the data after subtracting the best-fit profile. The expected centroid of H$\alpha$ is marked by a vertical blue line.
      }
    \label{fig:tr_Ha}
\end{figure*}

\begin{figure*}
   \centering
    \includegraphics[width=\linewidth,angle=0,trim={0cm 0cm 0cm 0cm},clip]{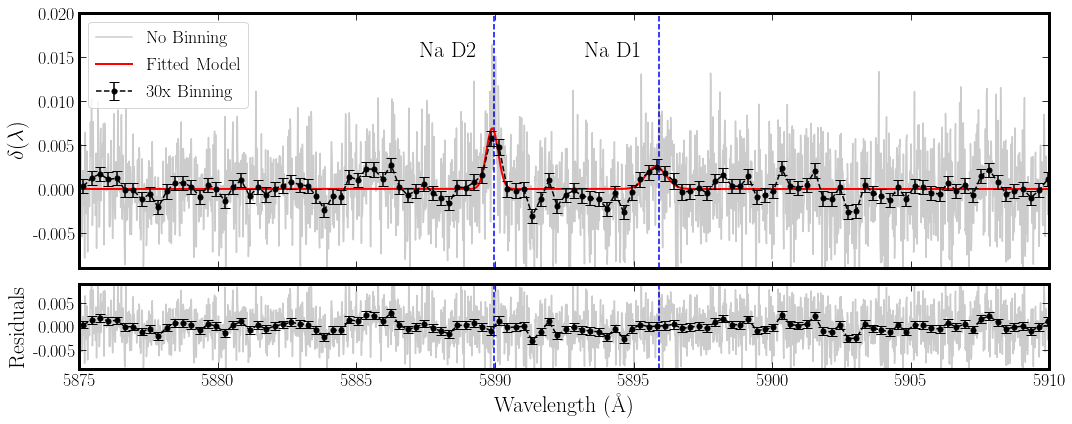}
      \caption{
      Same as Figure \ref{fig:tr_Ha}, for the Sodium D lines. The expected centroids of the D1 and D2 features are marked by vertical blue lines.
      }
    \label{fig:tr_Na}
\end{figure*}

\begin{figure*}
   \centering
   \includegraphics[width=\linewidth,angle=0,trim={0cm 0cm 0cm 0cm},clip]{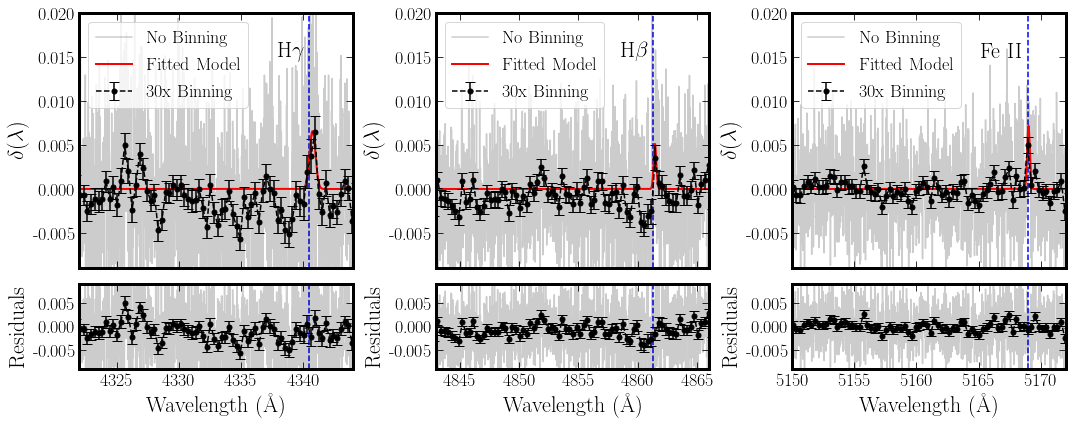}
      \caption{
      Same as Figure \ref{fig:tr_Ha}, for three potential atmospheric absorption features, H$\gamma$ ({\it Left}), H$\beta$ ({\it Middle}), Fe II ({\it Right}). The expected centroids are marked by vertical blue lines.
      }
    \label{fig:tr_extra}
\end{figure*}

We detect strong absorption from the atmosphere of WASP-121~b through multiple features in the transmission spectrum. We measure the line contrast, denoted $\mathcal{D}$ as the amplitude of a Gaussian profile fit to the absorption feature. We use the \texttt{Astropy} package, which performs a Levenberg-Marquardt Least-Squares fit. The H$\alpha$ line is detected with contrast $\mathcal{D} = 0.0187\pm0.0011$, centroid $\lambda_0 = 6562.93\pm0.02$ \AA, and $\text{FWHM} = 0.75 \pm 0.05$ \AA\ (Figure \ref{fig:tr_Ha}). We confirm the previous detection of the Sodium doublet \citep{Sindel2018}. We measure $\mathcal{D} = 0.0069\pm0.0012$, $\lambda_0 = 5890.01\pm0.06$ \AA, and $\text{FWHM} = 0.73 \pm 0.09$ \AA\ for the D2 line, and $\mathcal{D} = 0.0025\pm0.0009$, $\lambda_0 = 5896.09\pm0.09$ \AA, and $\text{FWHM} = 0.9 \pm 0.1$ \AA\ for the D1 line (Figure \ref{fig:tr_Na}). These measurements are comparable with those reported in \citep{Sindel2018}. The transmission spectra shown in Figures \ref{fig:tr_Ha} and \ref{fig:tr_Na} are in the rest-frame of the planet. The best fit centroids of the Na D lines are consistent with zero velocity offset. The H$\alpha$ line is offset by $+5.82\pm0.96$ \kms.
We scan the transmission spectrum for signs of additional features. We find nominally excess absorption at 4340.75 \AA, 4861.44 \AA, and 5169.02 \AA\ (Figure \ref{fig:tr_extra}), which we attribute to H$\gamma$ (4340.47 \AA), H$\beta$ (4861.33 \AA), and Fe II (5169.03 \AA) respectively. However, these features involve fewer data points and weaker line-contrasts. Hence we refrain from claiming definitive detections of these transitions.

\begin{figure*}
   \centering
    \includegraphics[width=\linewidth,angle=0,trim={0cm 0cm 0cm 0cm},clip]{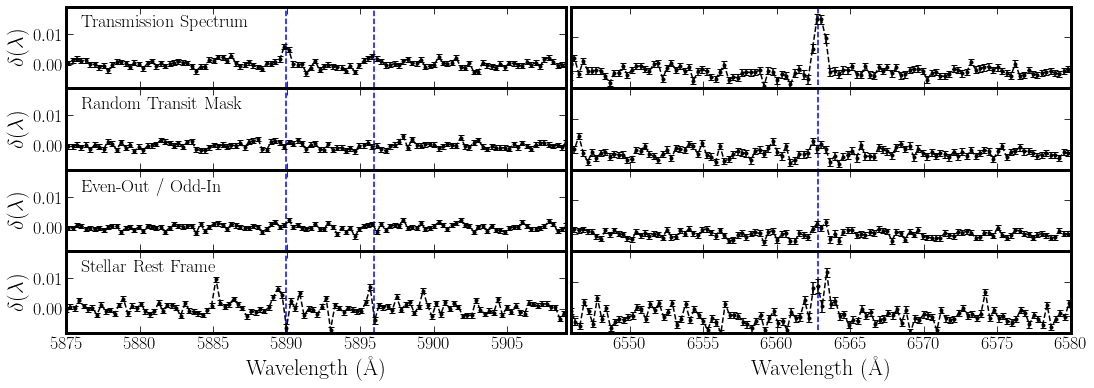}
      \caption{
      Excess transit depth from the atmosphere of WASP-121~b, under different test calculations of the transmission spectrum, for regions surround the Na doublet ({\it Left}) and H$\alpha$ ({\it Right}). {\it Top row}: the actual excess depth analysed in this study (no modifications); {\it Second row}: excess depth having randomized the in-transit and out-of-transit samples. {\it Third row}: excess depth with the out-of-transit sample composed of even frames, and the in-transit sample of odd frames. {\it Bottom row}: excess depth with the determined transit labels, but stacked in the rest-frame of the star instead of the planet.
      }
    \label{fig:tr_test}
\end{figure*}

In order to help rule out that our detections are from systematic or spurious artifacts, we perform the following control tests on the transmission spectrum: 1) randomization of in-transit and out-of-transit labels for individual exposures; 2) labelling even-numbered exposures as out-of-transit and odd-numbered exposures as in-transit; and 3) stacking the individual transmission spectra in the stellar rest frame. The resultant transmission spectra in the regions around the Na doublet and H$\alpha$ are shown in Figure \ref{fig:tr_test}. Mixing in-transit and out-of-transit spectra eliminates the signal completely. The stellar rest-frame is dominated by RM artifacts, but also preserves some of the planetary signal. However, the features are weaker and less coherent.

We specify absorption depth as the ratio of two fluxes \citep{Casasayas2017}. The first is the mean flux of the stacked transmission spectrum $\hat{\mathfrak{R}}(\lambda)$, within a narrow passband centered on a feature; the second is the mean flux along the continuum, sampling points at longer and shorter wavelengths. We then subtract their ratio from unity. Passbands centered on the feature have sizes 0.188, 0.375, 0.75, 1.50, 3.0 \AA. For the Na doublet, we select fluxes in the continuum spanning 5872.89-5884.89 \AA\ and 5900.89-5912.89 \AA. For H$\alpha$, the continuum ranges are 6480.0-6492.0 \AA\ and 6633.0-6645.0 \AA. We additionally compute an Na doublet average by combining the fluxes sampled in both passbands. Our results are summarized in Table \ref{tab:depth}. The narrow passbands produce absorption depths comparable to the fitted line contrasts. 

\renewcommand{\arraystretch}{1.0}
\begin{table*}
  \centering
\begin{tabular}{l|c|c|c|c|c|c|c|c|c|c|c}
Feature & $\mathcal{D}\:[\%]$ & 
AD$^{0.188 \mathring{\rm A}}_{\hat {\mathfrak{R}}}$ & 
AD$^{0.375 \mathring{\rm A}}_{\hat {\mathfrak{R}}}$ & 
AD$^{0.75 \mathring{\rm A}}_{\hat {\mathfrak{R}}}$ & 
AD$^{1.5 \mathring{\rm A}}_{\hat {\mathfrak{R}}}$ & 
AD$^{3.0 \mathring{\rm A}}_{\hat {\mathfrak{R}}}$ & 
AD$^{0.75 \mathring{\rm A}}_{\rm LC}$ & 
AD$^{1.5 \mathring{\rm A}}_{\rm LC}$ & 
AD$^{3.0 \mathring{\rm A}}_{\rm LC}$ & \\
\hline
\hline
H$\alpha$ & $1.87\pm0.11$ & $1.80\pm0.17$ & $1.68\pm0.12$ & $1.38\pm0.09$ & $0.92\pm0.06$ & $0.39\pm0.04$ & $1.29\pm0.17$ & $0.97\pm0.11$ & $0.40\pm0.06$ \\
Na D2 & $0.69\pm0.12$ & $0.89\pm0.11$ & $0.64\pm0.08$ & $0.49\pm0.06$ & $0.28\pm0.04$ & $0.10\pm0.03$ &  -  &  -  &  -  \\
Na D1 & $0.25\pm0.09$ & $0.42\pm0.11$ & $0.26\pm0.08$ & $0.23\pm0.06$ & $0.18\pm0.04$ & $0.06\pm0.03$ &  -  &  -  &  -  \\
Na Avg. &  -  & $0.65\pm0.08$ & $0.45\pm0.06$ & $0.36\pm0.04$ & $0.23\pm0.03$ & $0.08\pm0.02$ & $0.46\pm0.05$ & $0.31\pm0.04$ & $0.18\pm0.03$ \\
\label{tab:depth}
  \end{tabular}
  \caption{Summary of detections of H$\alpha$ and the Sodium D lines. All values are given in percentages. Column 2: line contrast measured via fit of Gaussian profile to transmission absorption feature. Columns 3-7: absorption depths of features in the stacked transmission spectrum, for the labeled passbands. Columns 8-10: absorption depths of transmission lightcurves, measured via fit of a Gaussian profile, for the labeled passbands.}
\end{table*}

\subsection{Lightcurve Analysis}

\begin{figure*}
   \centering
    \includegraphics[width=\linewidth,angle=0,trim={0cm 0cm 0cm 0cm},clip]{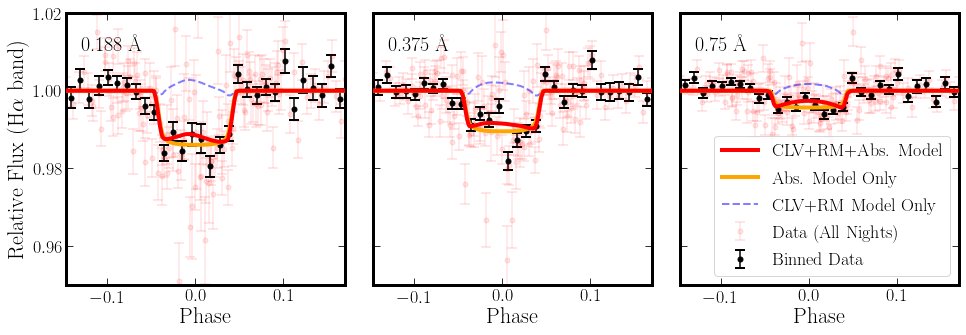}
      \caption{
      Transmission lightcurves for H$\alpha$ for 0.375, 0.75, and 1.50 \AA\ passbands. The blue dashed line depicts the theoretical contributions from center-to-limb variation and the Rossiter-McLaughlin effect. The red line depicts the best-fit absorption depth model, with included contributions from CLV and RM. Data from all nights were combined for the analysis.
      }
    \label{fig:lc_ha}
\end{figure*}

\begin{figure*}
   \centering
    \includegraphics[width=\linewidth,angle=0,trim={0cm 0cm 0cm 0cm},clip]{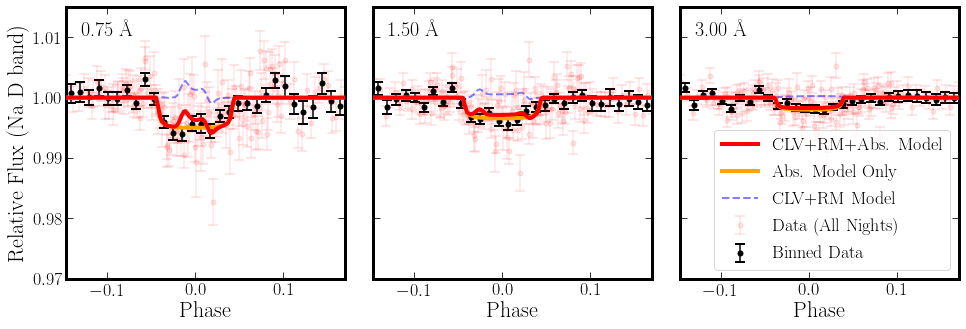}
      \caption{
      Same as Figure \ref{fig:lc_ha}, for the average depths of the Na D1 and Na D2 lines.
      }
    \label{fig:lc_na}
\end{figure*}

We compute photometric transit lightcurves in a similar manner as \citet{Casasayas2019}. We shift all transmission spectra $\mathfrak{R}(\lambda, t)$ to the planetary rest frame. For each spectrum, we determine the ratio of integrated line flux to integrated continuum flux, using the same 0.375, 0.75, 1.50 \AA\ passbands and continuum ranges as in the previous section. There is a small phase interval around mid-transit where CLV and RM effects overlap with atmospheric absorption, manifesting as sharp spikes in the transmission lightcurve. We model CLV and RM contributions by simulating the transmission lightcurve of a planet without an atmosphere at the same phases as the data (Section \ref{sec:clv}). We then divide the observed lightcurve by the model, and fit the residuals using the \texttt{PyTransit} package \citep{Parviainen2015a}. We estimate limb-darkening coefficients with the \texttt{LDTk} package \citep{Parviainen2015b}. The square of the fitted planet-to-star radii ratio is taken as an additional measurement of absorption depth. Lightcurves are shown in Figures \ref{fig:lc_ha} and \ref{fig:lc_na} for H$\alpha$ and the Na doublet average, respectively. Overall, the CLV+RM effect does not significantly impact the absorption depth measurement. The effect also averages out for the largest passbands. Minor differences between the observed and modelled CLV and RM effects may be attributed to our LTE assumption \citep{Casasayas2018}; also, we did not account for the effective radius of the planet $R_\lambda$, which is left as a free parameter in \citet{Casasayas2019}. The absorption depths are listed in Table \ref{tab:depth}, and are consistent with the stacked transmission spectrum absorption depths.

\subsection{Cross-Correlation with Atomic Species}

\begin{figure}
  \centering
   \includegraphics[width=\linewidth,angle=0,trim={0cm 0cm 0cm 0cm},clip]{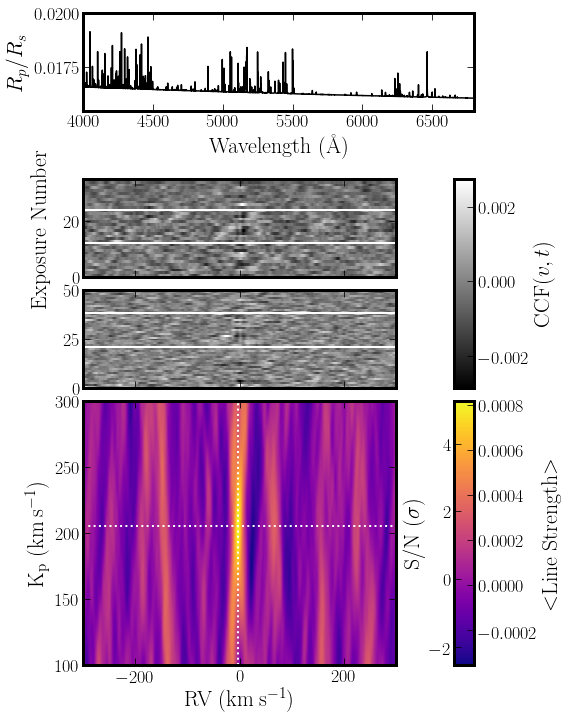}
      \caption{
      High-significance detection of Fe I via cross-correlation with a model template. {\it Top Panel}: Plot of the Fe I model spectrum, which was baseline-subtracted and normalized to obtain the cross-correlation template. {\it Middle Panel}: Time-series, residual cross-correlation functions (CCFs), which are obtained by subtracting the co-added out-of-transit CCF from all CCFs. The scaled Doppler Shadow model has also been subtracted, leaving the planet signal as a faint, light trail. Horizontal white lines mark the start and end of transit. Spectra were Doppler shifted by the known $V_{\rm sys}$ of WASP-121 prior to cross-correlation. The two rows correspond to Nights 1 and 3. {\it Bottom Panel}: Strength of the CCFs, co-added after being Doppler shifted by the velocity of WASP-121~b. Different assumptions are made for $K_p$ and $V_{\rm sys}$, which in turn sample different possible velocities of the planet. Dotted white lines mark the location of the peak signal. The color-map represents both detection significance, as well as the mean line-strength.
      }
    \label{fig:det_fei}
\end{figure}

We detect neutral Fe at $5.3\sigma$ significance. The cross-correlation signal peaks at $K_p = 205^{+30}_{-29}$ and $RV = -3^{+3}_{-1}$ \kms, where uncertainties correspond to 1$\sigma$ contours around the peak. Detection significance is defined as the number of standard deviations away from the mean in the entire sample of $K_p$-$V_{\rm sys}$ combinations (bottom panel, Figure \ref{fig:det_fei}). The $K_p$-$V_{\rm sys}$ samples from Nights 1 and 3 have been averaged together. Our quoted significance is a conservative estimate, since we refrain from sampling $K_p < 100$ \kms. We reduced the noise in this region when we subtracted the Doppler Shadow model, and its inclusion would decrease the overall sample variance. We also avoided optimizing weights and masking during cross-correlation, which could lead to spurious signals \citep{Cabot2019}. Considering the cross-correlation function as a weighted average line depth in the species, we may approximate an equivalent transit depth to be $0.00082\pm0.00014$. We note this is well below the noise levels of the stacked transmission spectrum, which demonstrates the advantage of cross-correlation for species with many lines.

\section{Discussion}
\label{sec:disc}

\subsection{Implications for Atmospheric Structure}

Based on line-contrasts, the H$\alpha$ line probes $R \sim 1.51 R_p$, and the Na D2 line probes $R \sim 1.20 R_p$. Our $\sim 2\%$ H$\alpha$ absorption depth suggests an extended Hydrogen atmosphere, possibly undergoing escape. \citet{Yan2018} discuss a similar scenario for KELT-9 b. The H$\alpha$ line is significantly redshifted, by $\sim6$ \kms. This measurement potentially probes high-velocity winds in the upper atmosphere of the planet. Winds were also reported in MASCARA-2 b by \citet{Casasayas2019}, who find a $-3.0$ to $-4.5$ \kms\ blueshifted H$\alpha$ line. A nominal estimate based on the average line strength of Fe I, as discussed in \S4.3, corresponds to $R \sim 1.03 R_p$. These heights represent optical depths of $\tau \sim 0.56$ for a chord tracing the atmosphere annulus around the planet \citep{Fortney2005, Etangs2008, Howe2012}. Fe I probably extends higher than $1.03 R_p$, since single transitions may have deeper transit depths than the weighted-average depth.

The recovered average Fe I line strength is dependent on both the weighting scheme and model template used in cross-correlation. Nevertheless, the species hardly extends out to the radii of Fe II as measured by \citet{Sing2019}, which reaches absorption depths of $\sim 10\%$. The scatter of their {\it HST} STIS NUV transmission spectrum reaches $R \sim 1.4 R_p$, which would prevent probing deep into the atmosphere. This explains why they only see one strong Fe I transition, corresponding to $R \sim 1.8 R_p$, whereas their ion detections reach nearly $R \sim 3 R_p$. Given the lower resolution at shorter wavelengths, cross-correlation may not be effective either. We note that Fe I opacity is strongest in the optical, and Fe II in the NUV. Hence, it follows that Fe is predominantly neutral at least out to $R \sim 1.03 R_p$, and possibly out to $\sim 1.4 R_p$.

Assuming $g = 843$ cm s$^{-2}$ (derived from Table \ref{tab:param}), $\mu = 2.22$ (mean molecular weight of Jupiter), and a nominal temperature of $T = T_{eq} = 2358$ K, WASP-121~b has a pressure scale height $H_p = 1039$ km. Note, WASP-121~b has a relatively large radius compared to other hot Jupiters, and is quite diffuse. Absorption out to one scale height produces an excess absorption depth of $\sim 240$ ppm, comparable to the value quoted by \citet{Evans2018}. Our Fe I absorption extends to $\sim 4 H_p$, which corresponds to a pressure of $\sim2\times10^{-3}$ bar assuming a typical $P_0 = 0.1$ bar corresponding to the white light radius \citep[e.g.][]{Welbanks2019}. The retrieved dayside Pressure-Temperature profile of WASP-121~b exhibits a thermal inversion \citep{Evans2017}, where the temperature rises steeply between $10^{-2} - 10^{-4}$ bar. Despite the suggestions of a thermal inversion in WASP-121b being caused by H-, TiO, and/or VO, it is possible that Fe absorption may also contribute to the same. Our calculations of thermochemical equilibrium show that Fe can remain largely neutral as high up in the atmosphere as 10$^{-6}$ bar for T $\sim$1500 - 3000 K, however photoionization is likely to contribute ionisation from Fe I to Fe II deeper in the atmosphere. 

\subsection{Center-to-Limb Variation \& Rossiter-McLaughlin Effects}
\label{sec:clv}

Ideally, dividing in-transit spectra by a co-added master out-of-transit spectrum isolates the atmospheric absorption. However, high-resolution spectroscopy resolves other differences between the out-of-transit spectrum and individual in-transit spectra. Center-to-Limb Variation describes the change in specific-intensity as a function of distance from the center of the star out to its limb \citep{Yan2017}, parametrized by the dimensionless quantity $1\geq \mu \geq 0$. Absorption lines in spectra from the limb of the star probe cooler gas at optical depth $\tau=1$. The continuum level also decreases from limb-darkening \citep{Mandel2002}. Another important effect is from stellar rotation, which Doppler shifts light emanating further from the rotation axis. An out-of-transit spectrum averages these effects over the entire stellar disk. However, during a transit, the planet occults a region of the disk with its own local stellar line profile, affected by CLV and rotation. This distorts the average stellar spectrum, and produces a residual in the ratio of in-transit and out-of-transit spectra (the distortion is called the Rossiter-McLauglin effect, only considering stellar rotation). The residual can create spurious features in the transmission spectrum and photometric lightcurves of individual lines \citep{Louden2015, Yan2015, Yan2017, Casasayas2018, Casasayas2019}.

\begin{figure*}
   \centering
    \includegraphics[width=1.0\linewidth,angle=0,trim={0cm 0cm 0cm 0cm},clip]{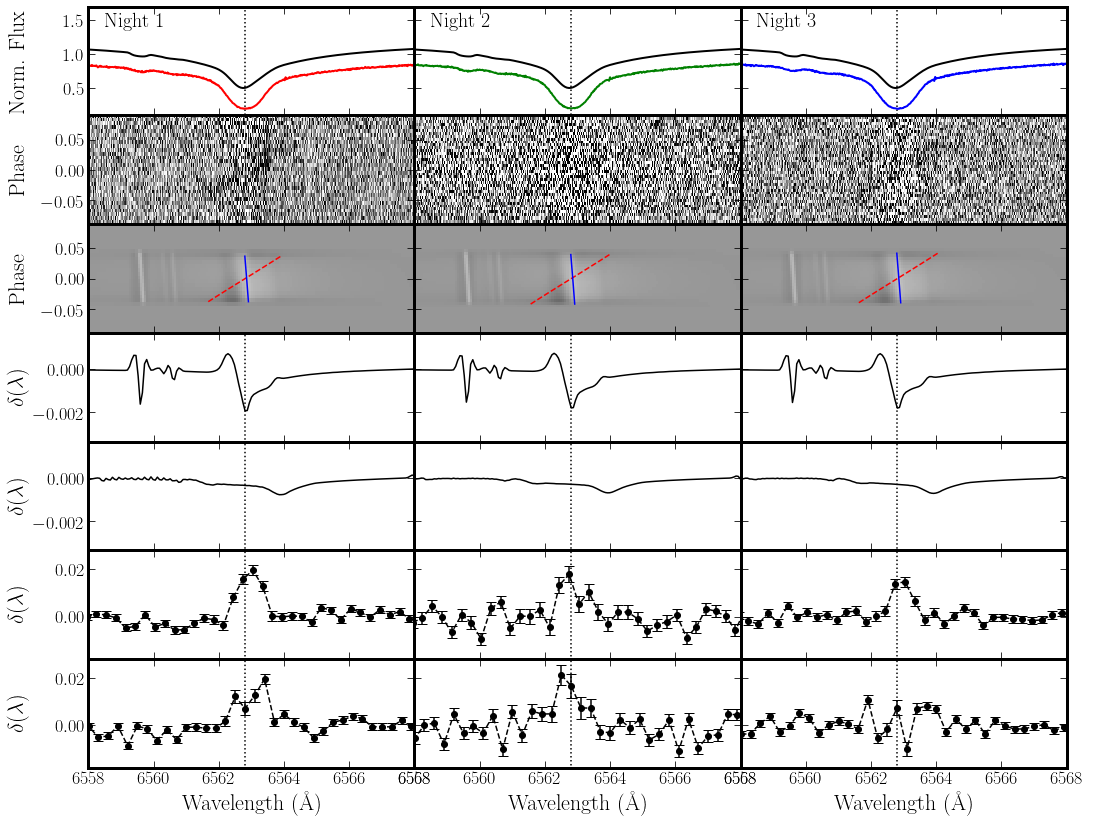}
      \caption{
      Comparison between the observed transmission spectra, and theoretical models of the CLV and RM effects, in the H$\alpha$ region. Each column represents a different night of observation. {\it Top Row}: the stacked and normalised out-of-transit stellar spectrum (colored line) and the integrated model stellar line profile (black line). {\it Second Row}: time-series transmission spectra for each night. The color range is from -0.05 (black) to 0.05 (white). Planetary atmosphere absorption is particularly visible in data from Nights 1 and 3. {\it Third Row}: modeled time-series transmission spectra, assuming the parameters in Table \ref{tab:param} and no atmosphere. The color range is from -0.01 (black) to 0.01 (white). The blue solid line denotes the position of H$\alpha$, Doppler shifted by the velocity of the occulted stellar region (RM effect). The red dashed line denotes the position of H$\alpha$, Doppler shifted by the expected planet velocity. {\it Fourth Row}: the jointly modeled CLV+RM effects, summed in the stellar rest-frame. {\it Fifth Row}: the jointly modeled CLV+RM effects, summed in the planetary rest-frame. {\it Sixth Row}: Transmission spectra summed in the planetary rest-frame. Note the change in y-axis scale. {\it Seventh Row}: Transmission spectra summed in the stellar rest-frame. Vertical dotted lines marked the rest-frame wavelength of H$\alpha$.
      }
    \label{fig:rmclv_ha}
\end{figure*}

\begin{figure*}
   \centering
    \includegraphics[width=1.0\linewidth,angle=0,trim={0cm 0cm 0cm 0cm},clip]{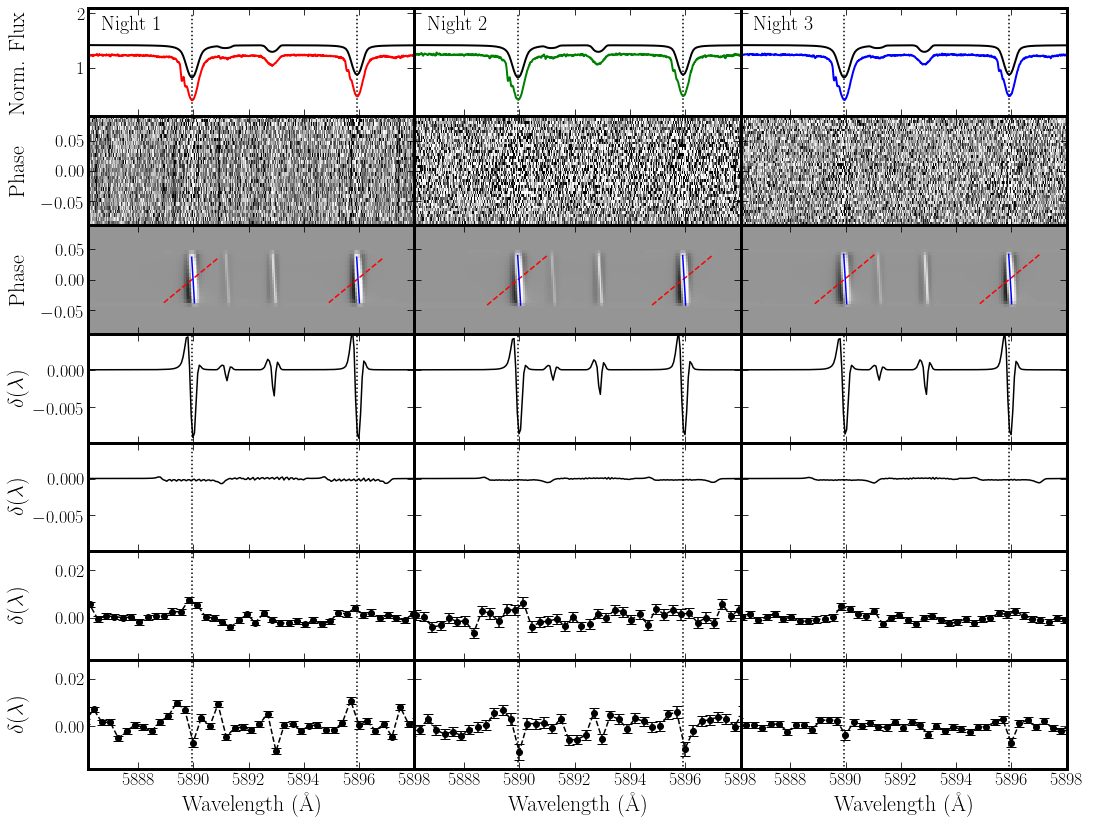}
      \caption{
      Same as Figure \ref{fig:rmclv_ha}, except for the Sodium D lines.
      }
    \label{fig:rmclv_na}
\end{figure*}

In order to evaluate the influence of CLV and RM effects in the observed data, we model them simultaneously in a manner similar to \citet{Casasayas2019}. We generate a synthetic spectrum of WASP-121 at 21 $\mu$-angles with \texttt{Spectroscopy Made Easy (SME)} \citep{Valenti1996}, using the VALD3 line-list database \citep{Ryabchikova2015} and Kurucz \texttt{ATLAS9} solar atmosphere models, and assuming parameters of $T_{eff} = 6460$ K, $\log g = 4.2$ and [Fe/H] = 0.13. We do not investigate non-LTE effects, or dependence of individual features on [Fe/H]. The stellar disk is simulated on an 80 $\times$ 80 pixel grid. Each pixel is allocated a spectrum, linearly interpolated between computed $\mu$ values, and Doppler shifted by the local rotation speed of the star. We assume $\lambda = -257.8^{\circ}$ and $v\sin i = 13.56$ \kms. We integrate the flux from each pixel over the full disk to obtain an out-of-transit spectrum. We model the transit according to the prescription of \citet{Cegla2016}, treating the planet as an opaque disk with no atmosphere. The planet's projected position on the stellar disk is given by,
\\
\begin{equation}
    x_p = \frac{a}{R_*}\sin{2\pi\phi}
\label{eqn:xp}
\end{equation}
\begin{equation}
    y_p = -\frac{a}{R_*}\cos{2\pi\phi}\cos{i_p}
\label{eqn:yp}
\end{equation}
\\
After rotating by the sky-projected obliquity, the position becomes,
\\
\begin{equation}
x_\perp = x_p\cos \lambda - y_p \sin \lambda
\end{equation}
\begin{equation}
y_\perp = x_p\sin \lambda + y_p \cos \lambda
\end{equation}
\\
and the planet occults a portion of the stellar disk with radial velocity,
\\
\begin{equation}
v_{\rm RM} = x_\perp v\sin i
\label{eqn:vrm}
\end{equation}
\\
The CLV is determined by the cosine of the angle ($\theta$) between the stellar normal and the observer,
\\
\begin{equation}
\mu = \cos\theta = (1 - (r_p/R_*)^2)^{1/2}
\label{eqn:mu}
\end{equation}
\\
where $r_p^2 = x_p^2 + y_p^2$ \citep{Mandel2002}. For each night, at each phase, we compute in-transit spectra by integrating over all pixels obscured by the planet, and subtracting the result from the out-of-transit spectrum. We then divide by the out-of-transit spectrum, normalize the continuum by a low-order polynomial, and subtract unity. For computational reasons, we narrow the wavelength range, around H$\alpha$ at first. We stack the transmission spectra in the rest frames of the planet and star. This model is shown in Figure \ref{fig:rmclv_ha}, along with the corresponding data. We repeat the analysis for the Na doublet region (Figure \ref{fig:rmclv_na}). The planet rest-frame residual is at the 0.073$\%$ level for H$\alpha$ and the 0.057$\%$ level for Na lines, which are negligible compared to the measured line strengths of 1.87\% for H$\alpha$ and 0.25-0.69\% for Na.  Further, they are less than the 1$\sigma$ uncertainties on the line depths, of 0.11\% for H$\alpha$ and 0.09-0.12\% for Na. If we increase the nominal radius of the planet by 1.5$\times$, then the residuals are at a 0.16\% level for H$\alpha$ and 0.12\%. While these are comparable to the uncertainties, the deepest parts of the residuals lie $\sim$1.5\AA\ redshifted from the line centroids, which further increases our confidence that the CLV and RM effects do not affect our absorption measurements. The stellar rest-frame residuals are about $3\times$ larger for H$\alpha$, and $15\times$ larger for the Na lines. 
In the third row of Figure \ref{fig:rmclv_ha}, we show the expected positions of H$\alpha$ features from planetary absorption (red-dashed line, corresponding to $v = v_{\rm pl}$) and the RM effect (blue-solid line, corresponding to $v = v_{\rm RM}$). Since the orbit of WASP-121~b is polar, these two trails are highly non-parallel. Hence, stacking in the planet frame (red-dashed line) smears out the already small RM effect (fifth row of Figure \ref{fig:rmclv_ha}). However, stacking in the stellar frame produces a large artifact, since the RM trail is nearly vertical (fourth row of Figure \ref{fig:rmclv_ha}). Note the stellar rest frame residuals are similar in shape and amplitude to the observed ones (fourth row, Figure \ref{fig:tr_test}), and the RM artifacts exhibit the negative-winged shape described in the previous section.

When do RM and CLV effects become significant in transmission spectroscopy? \citet{Yan2017} thoroughly explore CLV, finding that low $T_{eff}$ hosts exhibit stronger CLV artifacts. They also show the dependence on impact parameter ($b$). For the RM effect, one must also consider the sky projected obliquity ($\lambda$). We perform a systematic investigation of joint contributions of CLV and RM to transmission spectra for various orbits around WASP-121. As above, we model the planet as an opaque disk without an atmosphere in the following orbits: 1) the identical orbit as WASP-121~b, as modeled above; 2) an aligned orbit with $\lambda = 0.0$; 3) an exactly polar orbit of $\lambda = -270.0$; 4) an edge-on orbit with $b=0.0$; 5) an inclined orbit expected to maximize the CLV effect, $b=b_{\rm max}=0.84$ \citep{Yan2017}. Transmission spectra are calculated at 50 equally-spaced phases $-0.06 < \phi < 0.06$, and subsequently stacked in seven different rest-frames. These include the stellar rest-frame and the planet rest-frame under different assumptions of semi-amplitude ($K_p$). We additionally define an "RM-frame", which is Doppler shifted by the velocity of the occulted region of the stellar disk ($v_{\rm RM}$) (Equation \ref{eqn:vrm}). We note that explicit exclusion of ingress and egress spectra did not change results appreciably. All results are predominantly from the RM effect. This was tested by running the simulations with $v\sin i = 0.0$ \kms, which produced much smaller CLV-only artifacts. We did not explore dependence on the planet radius. A larger radius should exacerbate the effects \citep{DiGloria2015}. 

\begin{figure*}
   \centering
    \includegraphics[width=1.0\linewidth,angle=0,trim={0cm 0cm 0cm 0cm},clip]{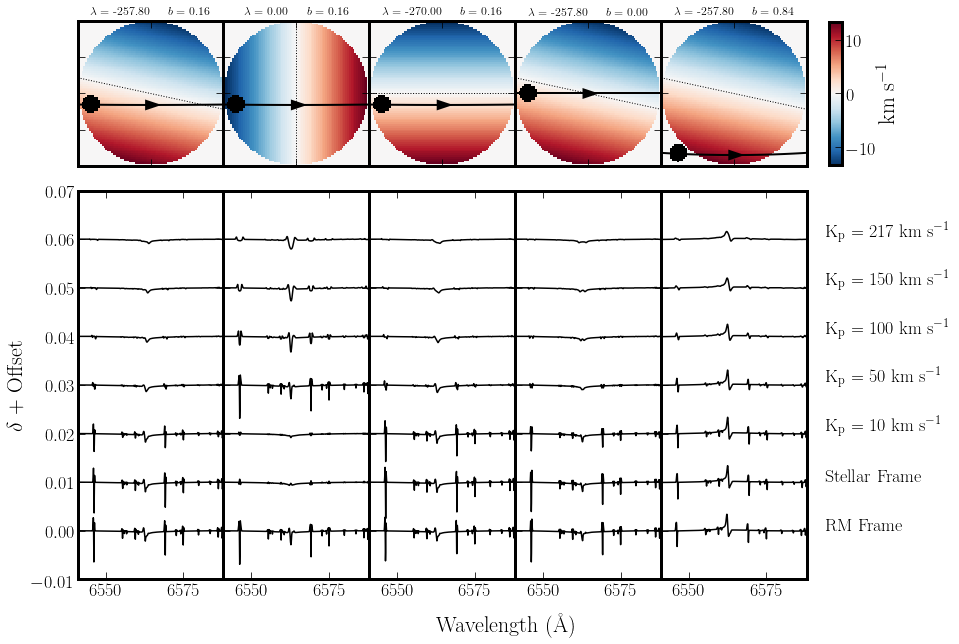}
      \caption{
      Theoretical contributions to transmission spectra from the jointly-modelled Center-to-Limb Variation and Rossiter-McLaughlin effects. We assume a star with effective temperature and radius of WASP-121, and a planet with the radius and semi-major axis of WASP-121~b. {\it Top Row}: planetary orbits explored for different choices of projected obliquity and impact parameter (the left-most column represents the actual orbit of WASP-121~b). The transmission spectra are stacked in seven different rest frames including: the velocites of the occulted stellar regions (RM Frame); zero-velocity (the Stellar Frame); and different assumptions of the planetary semi-amplitude, including the physical value for WASP-121~b of $217$ \kms. The analysis is restricted to the region around H$\alpha$.
      }
     \label{fig:rmclv_ha_multi}
\end{figure*}

Our results are shown in Figure \ref{fig:rmclv_ha_multi}. In nearly all cases, a growing $K_p$ increasingly smears out the artifacts. Setting $K_p = 10$ \kms\ corresponds to a slow, distant orbit, while $K_p = 217$ \kms\ is the fast, close-in orbit of WASP-121~b. The polar and near-polar orbits produce strong residuals in the stellar rest frame. As expected for an aligned orbit (second column), the RM effect approximately cancels out in the stellar rest frame, but is strong in the planet rest frame \citep{Louden2015}. In this case, it is notable that the artifacts spike at $K_p = 50$ \kms. This is because the planet and RM velocities are very similar (the trails are nearly parallel). As such, stacking in the planet frame is nearly equivalent to perfectly summing the RM artifact in its own frame (bottom row). This is an important consideration for observations of long-period planets. For a circular and edge-on orbit, one can estimate $K_p \simeq 2\pi a / P$, ingress and egress phases as $\phi_{\rm out} \simeq -\phi_{\rm in} \simeq R_*/2\pi a$, with $\Delta v_p = K_p(\sin2\pi\phi_{\rm out} - \sin2\pi\phi_{\rm in})$, and $\Delta v_{\rm RM} =  v_{\rm RM}(\phi_{\rm out}) - v_{\rm RM}(\phi_{\rm in})$ (which equals $v\sin i(\sin2\pi\phi_{\rm out} - \sin2\pi\phi_{\rm in})$ for an aligned orbit). RM artifacts are maximized when $\Delta v_p \sim \Delta v_{\rm RM}$.

\section{Conclusion}

The ultra-hot Jupiter WASP-121~b exhibits some of the most extreme and fascinating properties amongst giant exoplanets, including a particularly close-in orbit and high equilibrium temperature. It represents a corner case in planetary formation theory based on its orbital properties, and an important atmospheric case study for its thermal inversion \citep{Evans2017} and extended atmosphere \citep{Sing2019}. In this study, we have presented a detailed analysis of the optical transmission spectrum of WASP-121~b at high-resolution. We resolve nearly 2$\%$ excess atmospheric absorption from H$\alpha$, and 0.5-1$\%$ absorption from the Sodium D lines. The H$\alpha$ detection supports an extended and possibly escaping atmosphere. We present the additional high-significance detection of neutral Fe via the cross-correlation method \citep{Snellen2010}. While it is difficult to determine the exact extent of the region in the atmosphere containing Fe I, we find it lies approximately within stratosphere, possibly linking it to the thermal inversion. We additionally present a detailed analysis of the Rossiter-McLaughlin effect and Center-to-Limb variation, and how they impact the observed transmission spectra.

Our characterization of WASP-121~b comes at a time of systematic investigation of exoplanet atmospheres. Not only are multiple molecular and atomic detections being made in individual planets, but it is becoming possible to start comparing the chemical abundances of different planets \citep{Madhusudhan2019}. The detections of H, Na and Fe presented here add to a constantly growing list of detections in different wavelength regimes and at different spectral resolutions. Current and forthcoming generations of high-resolution optical spectrographs, including HARPS, HARPS-N, and the recently commissioned ESPRESSO \citep{Pepe2013} and EXPRES \citep{Jurgenson2016}, offer great potential for such studies, in pursuit of characterizing new planets, and in search of new and insightful chemistry.

\section*{Acknowledgements}
LW acknowledges support from the Gates Cambridge Trust towards his doctoral research. AAAP acknowledges support from STFC towards her doctoral research. SG acknowledges support from the UK Science and Technology Facilities Council (STFC) research grant ST/S000631/1. 

\bibliographystyle{mnras} 
\bibliography{main} 

\begin{thebibliography}{}
\makeatletter
\relax
\def\mn@urlcharsother{\let\do\@makeother \do\$\do\&\do\#\do\^\do\_\do\%\do\~}
\def\mn@doi{\begingroup\mn@urlcharsother \@ifnextchar [ {\mn@doi@}
  {\mn@doi@[]}}
\def\mn@doi@[#1]#2{\def\@tempa{#1}\ifx\@tempa\@empty \href
  {http://dx.doi.org/#2} {doi:#2}\else \href {http://dx.doi.org/#2} {#1}\fi
  \endgroup}
\def\mn@eprint#1#2{\mn@eprint@#1:#2::\@nil}
\def\mn@eprint@arXiv#1{\href {http://arxiv.org/abs/#1} {{\tt arXiv:#1}}}
\def\mn@eprint@dblp#1{\href {http://dblp.uni-trier.de/rec/bibtex/#1.xml}
  {dblp:#1}}
\def\mn@eprint@#1:#2:#3:#4\@nil{\def\@tempa {#1}\def\@tempb {#2}\def\@tempc
  {#3}\ifx \@tempc \@empty \let \@tempc \@tempb \let \@tempb \@tempa \fi \ifx
  \@tempb \@empty \def\@tempb {arXiv}\fi \@ifundefined
  {mn@eprint@\@tempb}{\@tempb:\@tempc}{\expandafter \expandafter \csname
  mn@eprint@\@tempb\endcsname \expandafter{\@tempc}}}

\bibitem[\protect\citeauthoryear{{Allart}, {Lovis}, {Pino}, {Wyttenbach},
  {Ehrenreich}  \& {Pepe}}{{Allart} et~al.}{2017}]{Allart2017}
{Allart} R.,  {Lovis} C.,  {Pino} L.,  {Wyttenbach} A.,  {Ehrenreich} D.,
  {Pepe} F.,  2017, \mn@doi [\aap] {10.1051/0004-6361/201730814}, \href
  {https://ui.adsabs.harvard.edu/abs/2017A%26A...606A.144A} {606, A144}

\bibitem[\protect\citeauthoryear{{Alonso-Floriano} et~al.,}{{Alonso-Floriano}
  et~al.}{2019}]{Alonso-Floriano2019}
{Alonso-Floriano} F.~J.,  et~al., 2019, \mn@doi [\aap]
  {10.1051/0004-6361/201834339}, \href
  {https://ui.adsabs.harvard.edu/abs/2019A&A...621A..74A} {621, A74}

\bibitem[\protect\citeauthoryear{{Anderson} et~al.,}{{Anderson}
  et~al.}{2018}]{Anderson2018}
{Anderson} D.~R.,  et~al., 2018, arXiv e-prints, \href
  {https://ui.adsabs.harvard.edu/abs/2018arXiv180904897A} {p. arXiv:1809.04897}

\bibitem[\protect\citeauthoryear{{Arcangeli} et~al.,}{{Arcangeli}
  et~al.}{2018}]{Arcangeli2018}
{Arcangeli} J.,  et~al., 2018, \mn@doi [\apjl] {10.3847/2041-8213/aab272},
  \href {https://ui.adsabs.harvard.edu/abs/2018ApJ...855L..30A} {855, L30}

\bibitem[\protect\citeauthoryear{{Asplund}, {Grevesse}, {Sauval}  \&
  {Scott}}{{Asplund} et~al.}{2009}]{Asplund2009}
{Asplund} M.,  {Grevesse} N.,  {Sauval} A.~J.,   {Scott} P.,  2009, \mn@doi
  [\araa] {10.1146/annurev.astro.46.060407.145222}, \href
  {https://ui.adsabs.harvard.edu/abs/2009ARA&A..47..481A} {47, 481}

\bibitem[\protect\citeauthoryear{{Batygin}, {Bodenheimer}  \&
  {Laughlin}}{{Batygin} et~al.}{2016}]{Batygin2016}
{Batygin} K.,  {Bodenheimer} P.~H.,   {Laughlin} G.~P.,  2016, \mn@doi [\apj]
  {10.3847/0004-637X/829/2/114}, \href
  {https://ui.adsabs.harvard.edu/abs/2016ApJ...829..114B} {829, 114}

\bibitem[\protect\citeauthoryear{Birkby}{Birkby}{2018}]{birkby2018}
Birkby J.~L.,  2018, Spectroscopic Direct Detection of Exoplanets.
Springer International Publishing, Cham, pp 1485--1508,
  \mn@doi{10.1007/978-3-319-55333-7_16}

\bibitem[\protect\citeauthoryear{{Birkby}, {de Kok}, {Brogi}, {de Mooij},
  {Schwarz}, {Albrecht}  \& {Snellen}}{{Birkby} et~al.}{2013}]{Birkby2013}
{Birkby} J.~L.,  {de Kok} R.~J.,  {Brogi} M.,  {de Mooij} E.~J.~W.,  {Schwarz}
  H.,  {Albrecht} S.,   {Snellen} I.~A.~G.,  2013, \mn@doi [\mnras]
  {10.1093/mnrasl/slt107}, \href
  {https://ui.adsabs.harvard.edu/abs/2013MNRAS.436L..35B} {436, L35}

\bibitem[\protect\citeauthoryear{{Birkby}, {de Kok}, {Brogi}, {Schwarz}  \&
  {Snellen}}{{Birkby} et~al.}{2017}]{Birkby2017}
{Birkby} J.~L.,  {de Kok} R.~J.,  {Brogi} M.,  {Schwarz} H.,   {Snellen}
  I.~A.~G.,  2017, \mn@doi [\aj] {10.3847/1538-3881/aa5c87}, \href
  {https://ui.adsabs.harvard.edu/abs/2017AJ....153..138B} {153, 138}

\bibitem[\protect\citeauthoryear{{Bond}, {Lauretta}  \& {O'Brien}}{{Bond}
  et~al.}{2010}]{bond2010}
{Bond} J.~C.,  {Lauretta} D.~S.,   {O'Brien} D.~P.,  2010, \mn@doi [\icarus]
  {10.1016/j.icarus.2009.07.037}, \href
  {http://adsabs.harvard.edu/abs/2010Icar..205..321B} {205, 321}

\bibitem[\protect\citeauthoryear{{Bourrier} et~al.,}{{Bourrier}
  et~al.}{2019}]{Bourrier2019}
{Bourrier} V.,  et~al., 2019, arXiv e-prints, \href
  {https://ui.adsabs.harvard.edu/abs/2019arXiv190903010B} {p. arXiv:1909.03010}

\bibitem[\protect\citeauthoryear{{Brogi}, {Snellen}, {de Kok}, {Albrecht},
  {Birkby}  \& {de Mooij}}{{Brogi} et~al.}{2012}]{Brogi2012}
{Brogi} M.,  {Snellen} I. A.~G.,  {de Kok} R.~J.,  {Albrecht} S.,  {Birkby} J.,
    {de Mooij} E. J.~W.,  2012, \mn@doi [\nat] {10.1038/nature11161}, \href
  {https://ui.adsabs.harvard.edu/abs/2012Natur.486..502B} {486, 502}

\bibitem[\protect\citeauthoryear{{Brogi}, {de Kok}, {Albrecht}, {Snellen},
  {Birkby}  \& {Schwarz}}{{Brogi} et~al.}{2016}]{Brogi2016}
{Brogi} M.,  {de Kok} R.~J.,  {Albrecht} S.,  {Snellen} I.~A.~G.,  {Birkby}
  J.~L.,   {Schwarz} H.,  2016, \mn@doi [\apj] {10.3847/0004-637X/817/2/106},
  \href {https://ui.adsabs.harvard.edu/abs/2016ApJ...817..106B} {817, 106}

\bibitem[\protect\citeauthoryear{{Brogi}, {Giacobbe}, {Guilluy}, {de Kok},
  {Sozzetti}, {Mancini}  \& {Bonomo}}{{Brogi} et~al.}{2018}]{brogi2018}
{Brogi} M.,  {Giacobbe} P.,  {Guilluy} G.,  {de Kok} R.~J.,  {Sozzetti} A.,
  {Mancini} L.,   {Bonomo} A.~S.,  2018, \mn@doi [\aap]
  {10.1051/0004-6361/201732189}, \href
  {https://ui.adsabs.harvard.edu/abs/2018A&A...615A..16B} {615, A16}

\bibitem[\protect\citeauthoryear{{Cabot}, {Madhusudhan}, {Hawker}  \&
  {Gandhi}}{{Cabot} et~al.}{2019}]{Cabot2019}
{Cabot} S. H.~C.,  {Madhusudhan} N.,  {Hawker} G.~A.,   {Gandhi} S.,  2019,
  \mn@doi [\mnras] {10.1093/mnras/sty2994}, \href
  {https://ui.adsabs.harvard.edu/abs/2019MNRAS.482.4422C} {482, 4422}

\bibitem[\protect\citeauthoryear{{Cartier} et~al.,}{{Cartier}
  et~al.}{2017}]{Cartier2017}
{Cartier} K. M.~S.,  et~al., 2017, \mn@doi [\aj] {10.3847/1538-3881/153/1/34},
  \href {https://ui.adsabs.harvard.edu/abs/2017AJ....153...34C} {153, 34}

\bibitem[\protect\citeauthoryear{{Casasayas-Barris}, {Palle}, {Nowak}, {Yan},
  {Nortmann}  \& {Murgas}}{{Casasayas-Barris} et~al.}{2017}]{Casasayas2017}
{Casasayas-Barris} N.,  {Palle} E.,  {Nowak} G.,  {Yan} F.,  {Nortmann} L.,
  {Murgas} F.,  2017, \mn@doi [\aap] {10.1051/0004-6361/201731956}, \href
  {https://ui.adsabs.harvard.edu/abs/2017A&A...608A.135C} {608, A135}

\bibitem[\protect\citeauthoryear{{Casasayas-Barris} et~al.,}{{Casasayas-Barris}
  et~al.}{2018}]{Casasayas2018}
{Casasayas-Barris} N.,  et~al., 2018, \mn@doi [\aap]
  {10.1051/0004-6361/201832963}, \href
  {http://adsabs.harvard.edu/abs/2018A%26A...616A.151C} {616, A151}

\bibitem[\protect\citeauthoryear{{Casasayas-Barris} et~al.,}{{Casasayas-Barris}
  et~al.}{2019}]{Casasayas2019}
{Casasayas-Barris} N.,  et~al., 2019, \mn@doi [\aap]
  {10.1051/0004-6361/201935623}, \href
  {https://ui.adsabs.harvard.edu/abs/2019A&A...628A...9C} {628, A9}

\bibitem[\protect\citeauthoryear{{Cauley}, {Shkolnik}, {Ilyin}, {Strassmeier},
  {Redfield}  \& {Jensen}}{{Cauley} et~al.}{2019}]{Cauley2019}
{Cauley} P.~W.,  {Shkolnik} E.~L.,  {Ilyin} I.,  {Strassmeier} K.~G.,
  {Redfield} S.,   {Jensen} A.,  2019, \mn@doi [\aj]
  {10.3847/1538-3881/aaf725}, \href
  {https://ui.adsabs.harvard.edu/abs/2019AJ....157...69C} {157, 69}

\bibitem[\protect\citeauthoryear{{Cegla}, {Lovis}, {Bourrier}, {Beeck},
  {Watson}  \& {Pepe}}{{Cegla} et~al.}{2016}]{Cegla2016}
{Cegla} H.~M.,  {Lovis} C.,  {Bourrier} V.,  {Beeck} B.,  {Watson} C.~A.,
  {Pepe} F.,  2016, \mn@doi [\aap] {10.1051/0004-6361/201527794}, \href
  {http://adsabs.harvard.edu/abs/2016A%26A...588A.127C} {588, A127}

\bibitem[\protect\citeauthoryear{{Dawson} \& {Johnson}}{{Dawson} \&
  {Johnson}}{2018}]{Dawson2018}
{Dawson} R.~I.,  {Johnson} J.~A.,  2018, \mn@doi [\araa]
  {10.1146/annurev-astro-081817-051853}, \href
  {https://ui.adsabs.harvard.edu/abs/2018ARA&A..56..175D} {56, 175}

\bibitem[\protect\citeauthoryear{{Daylan} et~al.,}{{Daylan}
  et~al.}{2019}]{Daylan2019}
{Daylan} T.,  et~al., 2019, arXiv e-prints, \href
  {https://ui.adsabs.harvard.edu/abs/2019arXiv190903000D} {p. arXiv:1909.03000}

\bibitem[\protect\citeauthoryear{{Delrez} et~al.,}{{Delrez}
  et~al.}{2016}]{Delrez2016}
{Delrez} L.,  et~al., 2016, \mn@doi [\mnras] {10.1093/mnras/stw522}, \href
  {https://ui.adsabs.harvard.edu/abs/2016MNRAS.458.4025D} {458, 4025}

\bibitem[\protect\citeauthoryear{{Di Gloria}, {Snellen}  \& {Albrecht}}{{Di
  Gloria} et~al.}{2015}]{DiGloria2015}
{Di Gloria} E.,  {Snellen} I.~A.~G.,   {Albrecht} S.,  2015, \mn@doi [\aap]
  {10.1051/0004-6361/201526218}, \href
  {https://ui.adsabs.harvard.edu/abs/2015A&A...580A..84D} {580, A84}

\bibitem[\protect\citeauthoryear{{Ehrenreich} et~al.,}{{Ehrenreich}
  et~al.}{2015}]{Ehrenreich2015}
{Ehrenreich} D.,  et~al., 2015, \mn@doi [\nat] {10.1038/nature14501}, \href
  {https://ui.adsabs.harvard.edu/abs/2015Natur.522..459E} {522, 459}

\bibitem[\protect\citeauthoryear{{Elser}, {Meyer}  \& {Moore}}{{Elser}
  et~al.}{2012}]{elser2012}
{Elser} S.,  {Meyer} M.~R.,   {Moore} B.,  2012, \mn@doi [\icarus]
  {10.1016/j.icarus.2012.09.016}, \href
  {http://adsabs.harvard.edu/abs/2012Icar..221..859E} {221, 859}

\bibitem[\protect\citeauthoryear{{Espinoza} et~al.,}{{Espinoza}
  et~al.}{2019}]{Espinoza2019}
{Espinoza} N.,  et~al., 2019, \mn@doi [\mnras] {10.1093/mnras/sty2691}, \href
  {https://ui.adsabs.harvard.edu/abs/2019MNRAS.482.2065E} {482, 2065}

\bibitem[\protect\citeauthoryear{{Evans} et~al.,}{{Evans}
  et~al.}{2016}]{Evans2016}
{Evans} T.~M.,  et~al., 2016, \mn@doi [\apjl] {10.3847/2041-8205/822/1/L4},
  \href {https://ui.adsabs.harvard.edu/abs/2016ApJ...822L...4E} {822, L4}

\bibitem[\protect\citeauthoryear{{Evans} et~al.,}{{Evans}
  et~al.}{2017}]{Evans2017}
{Evans} T.~M.,  et~al., 2017, \mn@doi [\nat] {10.1038/nature23266}, \href
  {https://ui.adsabs.harvard.edu/abs/2017Natur.548...58E} {548, 58}

\bibitem[\protect\citeauthoryear{{Evans} et~al.,}{{Evans}
  et~al.}{2018}]{Evans2018}
{Evans} T.~M.,  et~al., 2018, \mn@doi [\aj] {10.3847/1538-3881/aaebff}, \href
  {https://ui.adsabs.harvard.edu/abs/2018AJ....156..283E} {156, 283}

\bibitem[\protect\citeauthoryear{{Ford} \& {Rasio}}{{Ford} \&
  {Rasio}}{2006}]{Ford2006}
{Ford} E.~B.,  {Rasio} F.~A.,  2006, \mn@doi [\apjl] {10.1086/500734}, \href
  {https://ui.adsabs.harvard.edu/abs/2006ApJ...638L..45F} {638, L45}

\bibitem[\protect\citeauthoryear{{Fortney}}{{Fortney}}{2005}]{Fortney2005}
{Fortney} J.~J.,  2005, \mn@doi [\mnras] {10.1111/j.1365-2966.2005.09587.x},
  \href {https://ui.adsabs.harvard.edu/abs/2005MNRAS.364..649F} {364, 649}

\bibitem[\protect\citeauthoryear{{Fortney}, {Lodders}, {Marley}  \&
  {Freedman}}{{Fortney} et~al.}{2008}]{Fortney2008}
{Fortney} J.~J.,  {Lodders} K.,  {Marley} M.~S.,   {Freedman} R.~S.,  2008,
  \mn@doi [\apj] {10.1086/528370}, \href
  {https://ui.adsabs.harvard.edu/abs/2008ApJ...678.1419F} {678, 1419}

\bibitem[\protect\citeauthoryear{{Gandhi} \& {Madhusudhan}}{{Gandhi} \&
  {Madhusudhan}}{2017}]{gandhi_2017}
{Gandhi} S.,  {Madhusudhan} N.,  2017, \mn@doi [\mnras]
  {10.1093/mnras/stx1601}, \href
  {https://ui.adsabs.harvard.edu/abs/2017MNRAS.472.2334G} {472, 2334}

\bibitem[\protect\citeauthoryear{{Gandhi} \& {Madhusudhan}}{{Gandhi} \&
  {Madhusudhan}}{2019}]{Gandhi2019}
{Gandhi} S.,  {Madhusudhan} N.,  2019, \mn@doi [\mnras] {10.1093/mnras/stz751},
  \href {https://ui.adsabs.harvard.edu/abs/2019MNRAS.485.5817G} {485, 5817}

\bibitem[\protect\citeauthoryear{Gandhi, Madhusudhan, Hawker  \& Piette}{Gandhi
  et~al.}{2019}]{Gandhi2019b}
Gandhi S.,  Madhusudhan N.,  Hawker G.,   Piette A.,  2019, \mn@doi [\apj]
  {10.3847/1538-3881/ab4efc}, 158, 228

\bibitem[\protect\citeauthoryear{{Harrison}, {Bonsor}  \&
  {Madhusudhan}}{{Harrison} et~al.}{2018}]{harrison2018}
{Harrison} J.~H.~D.,  {Bonsor} A.,   {Madhusudhan} N.,  2018, \mn@doi [\mnras]
  {10.1093/mnras/sty1700}, \href
  {http://adsabs.harvard.edu/abs/2018MNRAS.479.3814H} {479, 3814}

\bibitem[\protect\citeauthoryear{{Hawker}, {Madhusudhan}, {Cabot}  \&
  {Gandhi}}{{Hawker} et~al.}{2018}]{Hawker2018}
{Hawker} G.~A.,  {Madhusudhan} N.,  {Cabot} S. H.~C.,   {Gandhi} S.,  2018,
  \mn@doi [\apj] {10.3847/2041-8213/aac49d}, \href
  {https://ui.adsabs.harvard.edu/abs/2018ApJ...863L..11H} {863, L11}

\bibitem[\protect\citeauthoryear{{Hoeijmakers} et~al.,}{{Hoeijmakers}
  et~al.}{2018}]{Hoeijmakers2018}
{Hoeijmakers} H.~J.,  et~al., 2018, \mn@doi [\nat] {10.1038/s41586-018-0401-y},
  \href {http://adsabs.harvard.edu/abs/2018Natur.560..453H} {560, 453}

\bibitem[\protect\citeauthoryear{{Hoeijmakers} et~al.,}{{Hoeijmakers}
  et~al.}{2019}]{Hoeijmakers2019}
{Hoeijmakers} H.~J.,  et~al., 2019, \mn@doi [\aap]
  {10.1051/0004-6361/201935089}, \href
  {https://ui.adsabs.harvard.edu/abs/2019A&A...627A.165H} {627, A165}

\bibitem[\protect\citeauthoryear{{Howe} \& {Burrows}}{{Howe} \&
  {Burrows}}{2012}]{Howe2012}
{Howe} A.~R.,  {Burrows} A.~S.,  2012, \mn@doi [\apj]
  {10.1088/0004-637X/756/2/176}, \href
  {https://ui.adsabs.harvard.edu/abs/2012ApJ...756..176H} {756, 176}

\bibitem[\protect\citeauthoryear{{Husser}, {Wende-von Berg}, {Dreizler},
  {Homeier}, {Reiners}, {Barman}  \& {Hauschildt}}{{Husser}
  et~al.}{2013}]{Husser2013}
{Husser} T.-O.,  {Wende-von Berg} S.,  {Dreizler} S.,  {Homeier} D.,  {Reiners}
  A.,  {Barman} T.,   {Hauschildt} P.~H.,  2013, \mn@doi [\aap]
  {10.1051/0004-6361/201219058}, \href
  {http://adsabs.harvard.edu/abs/2013A%26A...553A...6H} {553, A6}

\bibitem[\protect\citeauthoryear{{Jurgenson}, {Fischer}, {McCracken}, {Sawyer},
  {Szymkowiak}, {Davis}, {Muller}  \& {Santoro}}{{Jurgenson}
  et~al.}{2016}]{Jurgenson2016}
{Jurgenson} C.,  {Fischer} D.,  {McCracken} T.,  {Sawyer} D.,  {Szymkowiak} A.,
   {Davis} A.,  {Muller} G.,   {Santoro} F.,  2016, in \procspie. p. 99086T
  (\mn@eprint {arXiv} {1606.04413}), \mn@doi{10.1117/12.2233002}

\bibitem[\protect\citeauthoryear{{Kitzmann} et~al.,}{{Kitzmann}
  et~al.}{2018}]{Kitzmann2018}
{Kitzmann} D.,  et~al., 2018, \mn@doi [\apj] {10.3847/1538-4357/aace5a}, \href
  {https://ui.adsabs.harvard.edu/abs/2018ApJ...863..183K} {863, 183}

\bibitem[\protect\citeauthoryear{Kramida, {Yu.~Ralchenko}, Reader  \& {and NIST
  ASD Team}}{Kramida et~al.}{2018}]{Kramida2018}
Kramida A.,  {Yu.~Ralchenko} Reader J.,   {and NIST ASD Team} 2018, {NIST
  Atomic Spectra Database (ver. 5.6.1), [Online]. Available:
  {\tt{https://physics.nist.gov/asd}} [2019, February 6]. National Institute of
  Standards and Technology, Gaithersburg, MD.}

\bibitem[\protect\citeauthoryear{{Lecavelier Des Etangs}, {Pont},
  {Vidal-Madjar}  \& {Sing}}{{Lecavelier Des Etangs} et~al.}{2008}]{Etangs2008}
{Lecavelier Des Etangs} A.,  {Pont} F.,  {Vidal-Madjar} A.,   {Sing} D.,  2008,
  \mn@doi [\aap] {10.1051/0004-6361:200809388}, \href
  {https://ui.adsabs.harvard.edu/abs/2008A&A...481L..83L} {481, L83}

\bibitem[\protect\citeauthoryear{{Lockwood}, {Johnson}, {Bender}, {Carr},
  {Barman}, {Richert}  \& {Blake}}{{Lockwood} et~al.}{2014}]{lockwood2014}
{Lockwood} A.~C.,  {Johnson} J.~A.,  {Bender} C.~F.,  {Carr} J.~S.,  {Barman}
  T.,  {Richert} A.~J.~W.,   {Blake} G.~A.,  2014, \mn@doi [\apjl]
  {10.1088/2041-8205/783/2/L29}, \href
  {http://adsabs.harvard.edu/abs/2014ApJ...783L..29L} {783, L29}

\bibitem[\protect\citeauthoryear{{Lothringer}, {Barman}  \&
  {Koskinen}}{{Lothringer} et~al.}{2018}]{Lothringer2018}
{Lothringer} J.~D.,  {Barman} T.,   {Koskinen} T.,  2018, \mn@doi [\apj]
  {10.3847/1538-4357/aadd9e}, \href
  {https://ui.adsabs.harvard.edu/abs/2018ApJ...866...27L} {866, 27}

\bibitem[\protect\citeauthoryear{{Louden} \& {Wheatley}}{{Louden} \&
  {Wheatley}}{2015}]{Louden2015}
{Louden} T.,  {Wheatley} P.~J.,  2015, \mn@doi [\apjl]
  {10.1088/2041-8205/814/2/L24}, \href
  {https://ui.adsabs.harvard.edu/abs/2015ApJ...814L..24L} {814, L24}

\bibitem[\protect\citeauthoryear{{Madhusudhan}}{{Madhusudhan}}{2012}]{madhu2012}
{Madhusudhan} N.,  2012, \mn@doi [\apj] {10.1088/0004-637X/758/1/36}, \href
  {http://adsabs.harvard.edu/abs/2012ApJ...758...36M} {758, 36}

\bibitem[\protect\citeauthoryear{{Madhusudhan}}{{Madhusudhan}}{2019}]{Madhusudhan2019}
{Madhusudhan} N.,  2019, \mn@doi [\araa] {10.1146/annurev-astro-081817-051846},
  \href {https://ui.adsabs.harvard.edu/abs/2019ARA&A..57..617M} {57, 617}

\bibitem[\protect\citeauthoryear{{Madhusudhan}, {Knutson}, {Fortney}  \&
  {Barman}}{{Madhusudhan} et~al.}{2014a}]{madhu2014a}
{Madhusudhan} N.,  {Knutson} H.,  {Fortney} J.~J.,   {Barman} T.,  2014a, in
  {Beuther} H.,  {Klessen} R.~S.,  {Dullemond} C.~P.,   {Henning} T.,  eds,
  Protostars and Planets VI. p.~739 (\mn@eprint {arXiv} {1402.1169}),
  \mn@doi{10.2458/azu_uapress_9780816531240-ch032}

\bibitem[\protect\citeauthoryear{{Madhusudhan}, {Amin}  \&
  {Kennedy}}{{Madhusudhan} et~al.}{2014b}]{madhu2014c}
{Madhusudhan} N.,  {Amin} M.~A.,   {Kennedy} G.~M.,  2014b, \mn@doi [\apjl]
  {10.1088/2041-8205/794/1/L12}, \href
  {http://adsabs.harvard.edu/abs/2014ApJ...794L..12M} {794, L12}

\bibitem[\protect\citeauthoryear{{Mandel} \& {Agol}}{{Mandel} \&
  {Agol}}{2002}]{Mandel2002}
{Mandel} K.,  {Agol} E.,  2002, \mn@doi [\apjl] {10.1086/345520}, \href
  {https://ui.adsabs.harvard.edu/abs/2002ApJ...580L.171M} {580, L171}

\bibitem[\protect\citeauthoryear{{Mikal-Evans} et~al.,}{{Mikal-Evans}
  et~al.}{2019}]{Evans2019}
{Mikal-Evans} T.,  et~al., 2019, \mn@doi [\mnras] {10.1093/mnras/stz1753},
  \href {https://ui.adsabs.harvard.edu/abs/2019MNRAS.488.2222M} {488, 2222}

\bibitem[\protect\citeauthoryear{{Molli{\`e}re}, {van Boekel}, {Dullemond},
  {Henning}  \& {Mordasini}}{{Molli{\`e}re} et~al.}{2015}]{Molliere2015}
{Molli{\`e}re} P.,  {van Boekel} R.,  {Dullemond} C.,  {Henning} T.,
  {Mordasini} C.,  2015, \mn@doi [\apj] {10.1088/0004-637X/813/1/47}, \href
  {https://ui.adsabs.harvard.edu/abs/2015ApJ...813...47M} {813, 47}

\bibitem[\protect\citeauthoryear{{Moriarty}, {Madhusudhan}  \&
  {Fischer}}{{Moriarty} et~al.}{2014}]{moriarty2014}
{Moriarty} J.,  {Madhusudhan} N.,   {Fischer} D.,  2014, \mn@doi [\apj]
  {10.1088/0004-637X/787/1/81}, \href
  {http://adsabs.harvard.edu/abs/2014ApJ...787...81M} {787, 81}

\bibitem[\protect\citeauthoryear{{Nortmann} et~al.,}{{Nortmann}
  et~al.}{2018}]{Nortmann2018}
{Nortmann} L.,  et~al., 2018, \mn@doi [Science] {10.1126/science.aat5348},
  \href {https://ui.adsabs.harvard.edu/abs/2018Sci...362.1388N} {362, 1388}

\bibitem[\protect\citeauthoryear{{Nugroho}, {Kawahara}, {Masuda}, {Hirano},
  {Kotani}  \& {Tajitsu}}{{Nugroho} et~al.}{2017}]{Nugroho2017}
{Nugroho} S.~K.,  {Kawahara} H.,  {Masuda} K.,  {Hirano} T.,  {Kotani} T.,
  {Tajitsu} A.,  2017, \mn@doi [\aj] {10.3847/1538-3881/aa9433}, \href
  {https://ui.adsabs.harvard.edu/abs/2017AJ....154..221N} {154, 221}

\bibitem[\protect\citeauthoryear{{{\"O}berg}, {Murray-Clay}  \&
  {Bergin}}{{{\"O}berg} et~al.}{2011}]{oberg2011}
{{\"O}berg} K.~I.,  {Murray-Clay} R.,   {Bergin} E.~A.,  2011, \mn@doi [\apj]
  {10.1088/2041-8205/743/1/L16}, \href
  {http://adsabs.harvard.edu/abs/2011ApJ...743L..16O} {743, L16}

\bibitem[\protect\citeauthoryear{{Parmentier} et~al.,}{{Parmentier}
  et~al.}{2018}]{Parmentier2018}
{Parmentier} V.,  et~al., 2018, \mn@doi [\aap] {10.1051/0004-6361/201833059},
  \href {https://ui.adsabs.harvard.edu/abs/2018A&A...617A.110P} {617, A110}

\bibitem[\protect\citeauthoryear{Parviainen}{Parviainen}{2015}]{Parviainen2015a}
Parviainen H.,  2015, \mn@doi [MNRAS] {10.1093/mnras/stv894}, 450, 3233

\bibitem[\protect\citeauthoryear{Parviainen \& Aigrain}{Parviainen \&
  Aigrain}{2015}]{Parviainen2015b}
Parviainen H.,  Aigrain S.,  2015, \mn@doi [MNRAS] {10.1093/mnras/stv1857},
  453, 3821

\bibitem[\protect\citeauthoryear{{Pasek}, {Milsom}, {Ciesla}, {Lauretta},
  {Sharp}  \& {Lunine}}{{Pasek} et~al.}{2005}]{pasek2005}
{Pasek} M.~A.,  {Milsom} J.~A.,  {Ciesla} F.~J.,  {Lauretta} D.~S.,  {Sharp}
  C.~M.,   {Lunine} J.~I.,  2005, \mn@doi [\icarus]
  {10.1016/j.icarus.2004.10.012}, \href
  {http://adsabs.harvard.edu/abs/2005Icar..175....1P} {175, 1}

\bibitem[\protect\citeauthoryear{{Pepe} et~al.,}{{Pepe}
  et~al.}{2013}]{Pepe2013}
{Pepe} F.,  et~al., 2013, The Messenger, \href
  {http://adsabs.harvard.edu/abs/2013Msngr.153....6P} {153, 6}

\bibitem[\protect\citeauthoryear{{Pinhas}, {Rackham}, {Madhusudhan}  \&
  {Apai}}{{Pinhas} et~al.}{2018}]{Pinhas2018}
{Pinhas} A.,  {Rackham} B.~V.,  {Madhusudhan} N.,   {Apai} D.,  2018, \mn@doi
  [\mnras] {10.1093/mnras/sty2209}, \href
  {https://ui.adsabs.harvard.edu/abs/2018MNRAS.480.5314P} {480, 5314}

\bibitem[\protect\citeauthoryear{{Piskorz} et~al.,}{{Piskorz}
  et~al.}{2016}]{piskorz2016}
{Piskorz} D.,  et~al., 2016, \mn@doi [\apj] {10.3847/0004-637X/832/2/131},
  \href {http://adsabs.harvard.edu/abs/2016ApJ...832..131P} {832, 131}

\bibitem[\protect\citeauthoryear{{Richard} et~al.,}{{Richard}
  et~al.}{2012}]{Richard2012}
{Richard} C.,  et~al., 2012, \mn@doi [\jqsrt] {10.1016/j.jqsrt.2011.11.004},
  \href {http://adsabs.harvard.edu/abs/2012JQSRT.113.1276R} {113, 1276}

\bibitem[\protect\citeauthoryear{{Rodler}, {K{\"u}rster}  \& {Barnes}}{{Rodler}
  et~al.}{2013}]{rodler2013}
{Rodler} F.,  {K{\"u}rster} M.,   {Barnes} J.~R.,  2013, \mn@doi [\mnras]
  {10.1093/mnras/stt462}, \href
  {http://adsabs.harvard.edu/abs/2013MNRAS.432.1980R} {432, 1980}

\bibitem[\protect\citeauthoryear{{Ryabchikova}, {Piskunov}, {Kurucz},
  {Stempels}, {Heiter}, {Pakhomov}  \& {Barklem}}{{Ryabchikova}
  et~al.}{2015}]{Ryabchikova2015}
{Ryabchikova} T.,  {Piskunov} N.,  {Kurucz} R.~L.,  {Stempels} H.~C.,  {Heiter}
  U.,  {Pakhomov} Y.,   {Barklem} P.~S.,  2015, \mn@doi [\physscr]
  {10.1088/0031-8949/90/5/054005}, \href
  {https://ui.adsabs.harvard.edu/abs/2015PhyS...90e4005R} {90, 054005}

\bibitem[\protect\citeauthoryear{{Salz}, {Schneider}, {Fossati}, {Czesla},
  {France}  \& {Schmitt}}{{Salz} et~al.}{2019}]{Salz2019}
{Salz} M.,  {Schneider} P.~C.,  {Fossati} L.,  {Czesla} S.,  {France} K.,
  {Schmitt} J.~H.~M.~M.,  2019, \mn@doi [\aap] {10.1051/0004-6361/201732419},
  \href {https://ui.adsabs.harvard.edu/abs/2019A&A...623A..57S} {623, A57}

\bibitem[\protect\citeauthoryear{{Seidel} et~al.,}{{Seidel}
  et~al.}{2019}]{Seidel2019}
{Seidel} J.~V.,  et~al., 2019, \mn@doi [\aap] {10.1051/0004-6361/201834776},
  \href {https://ui.adsabs.harvard.edu/abs/2019A&A...623A.166S} {623, A166}

\bibitem[\protect\citeauthoryear{{Sheppard}, {Mandell}, {Tamburo}, {Gand hi},
  {Pinhas}, {Madhusudhan}  \& {Deming}}{{Sheppard} et~al.}{2017}]{Sheppard2017}
{Sheppard} K.~B.,  {Mandell} A.~M.,  {Tamburo} P.,  {Gand hi} S.,  {Pinhas} A.,
   {Madhusudhan} N.,   {Deming} D.,  2017, \mn@doi [\apjl]
  {10.3847/2041-8213/aa9ae9}, \href
  {https://ui.adsabs.harvard.edu/abs/2017ApJ...850L..32S} {850, L32}

\bibitem[\protect\citeauthoryear{Sindel}{Sindel}{2018}]{Sindel2018}
Sindel J.~P.,  2018, {Masters} dissertation, Lule${\rm \mathring{a}}$
  University

\bibitem[\protect\citeauthoryear{{Sing} et~al.,}{{Sing}
  et~al.}{2019}]{Sing2019}
{Sing} D.~K.,  et~al., 2019, \mn@doi [\aj] {10.3847/1538-3881/ab2986}, \href
  {https://ui.adsabs.harvard.edu/abs/2019AJ....158...91S} {158, 91}

\bibitem[\protect\citeauthoryear{{Smette} et~al.,}{{Smette}
  et~al.}{2015}]{smette2015}
{Smette} A.,  et~al., 2015, \mn@doi [\aap] {10.1051/0004-6361/201423932}, \href
  {https://ui.adsabs.harvard.edu/abs/2015A&A...576A..77S} {576, A77}

\bibitem[\protect\citeauthoryear{{Snellen}, {de Kok}, {de Mooij}  \&
  {Albrecht}}{{Snellen} et~al.}{2010}]{Snellen2010}
{Snellen} I. A.~G.,  {de Kok} R.~J.,  {de Mooij} E. J.~W.,   {Albrecht} S.,
  2010, \mn@doi [\nat] {10.1038/nature09111}, \href
  {https://ui.adsabs.harvard.edu/abs/2010Natur.465.1049S} {465, 1049}

\bibitem[\protect\citeauthoryear{{Tamuz}, {Mazeh}  \& {Zucker}}{{Tamuz}
  et~al.}{2005}]{tamuz_2005}
{Tamuz} O.,  {Mazeh} T.,   {Zucker} S.,  2005, \mn@doi [\mnras]
  {10.1111/j.1365-2966.2004.08585.x}, \href
  {http://adsabs.harvard.edu/abs/2005MNRAS.356.1466T} {356, 1466}

\bibitem[\protect\citeauthoryear{{Triaud} et~al.,}{{Triaud}
  et~al.}{2010}]{Triaud2010}
{Triaud} A.~H.~M.~J.,  et~al., 2010, \mn@doi [\aap]
  {10.1051/0004-6361/201014525}, \href
  {https://ui.adsabs.harvard.edu/abs/2010A&A...524A..25T} {524, A25}

\bibitem[\protect\citeauthoryear{{Valenti} \& {Piskunov}}{{Valenti} \&
  {Piskunov}}{1996}]{Valenti1996}
{Valenti} J.~A.,  {Piskunov} N.,  1996, \aaps, \href
  {https://ui.adsabs.harvard.edu/abs/1996A&AS..118..595V} {118, 595}

\bibitem[\protect\citeauthoryear{{Welbanks} \& {Madhusudhan}}{{Welbanks} \&
  {Madhusudhan}}{2019}]{Welbanks2019}
{Welbanks} L.,  {Madhusudhan} N.,  2019, \mn@doi [\aj]
  {10.3847/1538-3881/ab14de}, \href
  {https://ui.adsabs.harvard.edu/abs/2019AJ....157..206W} {157, 206}

\bibitem[\protect\citeauthoryear{{Wyttenbach}, {Ehrenreich}, {Lovis}, {Udry}
  \& {Pepe}}{{Wyttenbach} et~al.}{2015}]{Wyttenbach2015}
{Wyttenbach} A.,  {Ehrenreich} D.,  {Lovis} C.,  {Udry} S.,   {Pepe} F.,  2015,
  \mn@doi [\aap] {10.1051/0004-6361/201525729}, \href
  {https://ui.adsabs.harvard.edu/abs/2015A&A...577A..62W} {577, A62}

\bibitem[\protect\citeauthoryear{{Yan} \& {Henning}}{{Yan} \&
  {Henning}}{2018}]{Yan2018}
{Yan} F.,  {Henning} T.,  2018, \mn@doi [Nature Astronomy]
  {10.1038/s41550-018-0503-3}, \href
  {https://ui.adsabs.harvard.edu/abs/2018NatAs...2..714Y} {2, 714}

\bibitem[\protect\citeauthoryear{{Yan}, {Fosbury}, {Petr-Gotzens}, {Zhao}  \&
  {Pall{\'e}}}{{Yan} et~al.}{2015}]{Yan2015}
{Yan} F.,  {Fosbury} R.~A.~E.,  {Petr-Gotzens} M.~G.,  {Zhao} G.,   {Pall{\'e}}
  E.,  2015, \mn@doi [\aap] {10.1051/0004-6361/201425220}, \href
  {https://ui.adsabs.harvard.edu/abs/2015A&A...574A..94Y} {574, A94}

\bibitem[\protect\citeauthoryear{{Yan}, {Pall{\'e}}, {Fosbury}, {Petr-Gotzens}
  \& {Henning}}{{Yan} et~al.}{2017}]{Yan2017}
{Yan} F.,  {Pall{\'e}} E.,  {Fosbury} R.~A.~E.,  {Petr-Gotzens} M.~G.,
  {Henning} T.,  2017, \mn@doi [\aap] {10.1051/0004-6361/201630144}, \href
  {https://ui.adsabs.harvard.edu/abs/2017A%26A...603A..73Y} {603, A73}

\makeatother
\end{thebibliography}

\bsp	
\label{lastpage}
\end{document}